\pdfoutput=1
\documentclass[11pt,a4paper]{article}
\usepackage{jheppub}
\usepackage[caption=false]{subfig}
\usepackage{graphicx}
\usepackage{multirow}
\usepackage{amsbsy}
\usepackage{color}

\def\p{\partial}

\def\=:{=\hspace{-.7em}\raisebox{1.1ex}{.}\hspace{.1em}\raisebox{-0.2ex}{.}}
\newcommand{\Tr}{{\rm Tr}\,}

\newcommand {\beq}{\begin{eqnarray}}
\newcommand {\eeq}{\end{eqnarray}}
\newcommand {\non}{\nonumber\\}

\newcommand {\ab}[1]{\langle#1\rangle}
\newcommand {\diag}{\mathop{\rm diag}}
\newcommand {\llls}[3]{\lambda_1^{#1}\lambda_2^{#2}\lambda_3^{#3}}
\newcommand {\llij}[4]{\lambda_{#1}^{#3}\lambda_{#2}^{#4}}
\newcommand {\lllsbar}[3]{\bar{\lambda}_1^{#1}\bar{\lambda}_2^{#2}\bar{\lambda}_3^{#3}} 
\newcommand {\llijbar}[4]{\bar{\lambda}_{#1}^{#3}\bar{\lambda}_{#2}^{#4}}

\newtheorem{conjecture}{Conjecture}

\begin{document}

\title{A higher-order Skyrme model}

\author{Sven Bjarke Gudnason${}^1$,}
\author{Muneto Nitta${}^2$}
\affiliation{${}^1$Institute of Modern Physics, Chinese Academy of
  Sciences, Lanzhou 730000, China}
\affiliation{${}^2$Department of Physics, and Research and Education
  Center for Natural Sciences, Keio University, Hiyoshi 4-1-1,
  Yokohama, Kanagawa 223-8521, Japan}
\emailAdd{bjarke(at)impcas.ac.cn}
\emailAdd{nitta(at)phys-h.keio.ac.jp}

\abstract{
We propose a higher-order Skyrme model with derivative terms of
eighth, tenth and twelfth order.
Our construction yields simple and easy-to-interpret higher-order
Lagrangians.
We first show that a Skyrmion with higher-order terms proposed by
Marleau has an instability in the form of a baby-Skyrmion string, 
while the static energies of our construction are positive definite, 
implying stability against time-independent perturbations.
However, we also find that the Hamiltonians of our construction
possess two kinds of dynamical instabilities, which may indicate the
instability with respect to time-dependent perturbations.
Different from the well-known Ostrogradsky instability, 
the instabilities that we find are intrinsically of nonlinear nature  
and also due to the fact that even powers of the inverse metric gives
a ghost-like higher-order kinetic-like term.
The vacuum state is, however, stable. 
Finally, we show that at sufficiently low energies, our Hamiltonians
in the simplest cases, are stable against time-dependent
perturbations. 
}


\maketitle

\section{Introduction}

The Skyrme model \cite{Skyrme:1962vh,Skyrme:1961vq} is generally
believed to describe low-energy QCD at large
$N_c$ \cite{Witten:1983tw,Witten:1983tx}.
It has also been derived directly from the QCD Lagrangian by means of
partial bosonization \cite{Zaks:1985cv}. As well known, it is not
possible to perform a full bosonization in 3+1 dimensions and hence
the latter reference is bosonizing only the phases of the fermions.
The Skyrme model has also been derived in the Sakai-Sugimoto
model \cite{Sakai:2004cn} by considering the effective action for the
zero mode.
All these derivations of the Skyrme model include a kinetic term as
well as the \emph{Skyrme} term, which is fourth order in derivatives.
Skyrme introduced the term \cite{Skyrme:1962vh,Skyrme:1961vq} in order
to stabilize the soliton -- the Skyrmion -- from collapse, as
otherwise is unavoidable due to Derrick's
theorem \cite{Derrick:1964ww}. 
However, higher-order derivative corrections higher than fourth order
are generally expected.

As expected in QCD and explicitly shown in the Sakai-Sugimoto
model \cite{Sakai:2004cn}, an infinite tower of vector mesons exist as
one goes up in energy scales.
For each of these massive vector mesons, one can obtain effective
operators in a pure pion theory by integrating out the massive
mesons. The interaction terms between the pions and the mesons yield
new low-energy effective operators.
The first higher-derivative correction to the Skyrme model is expected
to be a sixth-order derivative term, see
e.g.~\cite{Adkins:1983nw,Jackson:1985yz,Marleau:1989fh,Marleau:1990nh,Marleau:1991jk,Marleau:2000ia,Adam:2010fg,Adam:2010ds,Gudnason:2013qba,Gudnason:2014jga,Gudnason:2014hsa,Gudnason:2015nxa,Gudnason:2016yix}.
Physically, it corresponds to integrating out the
$\omega$-meson \cite{Adkins:1983nw,Jackson:1985yz}; this can be seen
from the phenomenological Lagrangian with the interaction describing
the decay $\omega\to\pi^+\pi^-\pi^0$.

The sixth-order term -- which we shall call the BPS-Skyrme term --
recently caught interest due to its BPS properties when it is paired
with a suitable potential \cite{Adam:2010fg,Adam:2010ds,Gudnason:2013qba,Gudnason:2014jga,Gudnason:2014hsa,Gudnason:2015nxa,Gudnason:2016yix}. Here BPS
simply means that the energy is proportional to the topological number
-- the Skyrmion number, $B$ -- of the model.\footnote{For
supersymmetrizations of the Skyrme model, see
e.g.~Refs.~\cite{Bergshoeff:1984wb,Freyhult:2003zb,Queiruga:2015xka,Gudnason:2015ryh}. } 
This is a desired feature in nuclear physics where binding energies
are very small. 

In principle, we expect infinitely many higher-derivative terms in the
low-energy effective action. However, as each term is larger in
canonical dimension, it necessarily has to be accompanied by a
dimensional constant to the same power minus four.
That constant is typically proportional to the mass of the state that
was integrated out of the underlying theory.
Therefore, as long as the energy scales being probed are much smaller
than the lowest mass scale of a state that was integrated out, the
higher-derivative expansion may make sense and thus
converge\footnote{Mathematically, such series may not be well-defined
or converge in any mathematical sense. We will not dwell upon such
obstacles here. }.

Apart from a construction based on the hedgehog Ansatz by
Marleau \cite{Marleau:1989fh,Marleau:1990nh,Marleau:1991jk,Marleau:2000ia},
no extensive studies on higher-derivative terms in 3+1 dimensions,
higher than sixth order, has been carried out in the
literature\footnote{Ref.~\cite{Nakamula:2016wwv} considered a 
higher-dimensional generalization of the Atiyah-Manton construction of
Skyrm\-ions using the holonomy of instantons; this reference considers
eighth-order derivative terms in 7+1 dimensions. } -- to the best of
our knowledge.
Marleau considered a construction that yielded higher-order derivative
corrections to the Skyrmion, but restricted in such a way as to give
only a second-order equation of motion for the radial profile (chiral
angle
function) \cite{Marleau:1989fh,Marleau:1990nh,Marleau:1991jk,Marleau:2000ia}. 
When restricted to spherical symmetry, this construction gives stable 
profiles when certain stability criteria are
satisfied \cite{Jackson:1991mt}. 
Nevertheless, as we will show in Sec.~\ref{sec:Marleau}, when relaxing
the spherical symmetry, this construction becomes unstable.
Longpr\'e and Marleau later found that avoiding the instability was
indeed difficult \cite{Longpre:2005fr,Longpre:2007zz}; they proposed a
stability criterion that, however, cannot be satisfied for a
finite-order derivative Lagrangian without causing Derrick
instability. 
We will propose our interpretation of the instability as well as why
it occurs and show that to finite order, it cannot be cured
(stabilized). 
The instability occurs if perturbations are independent of one spatial 
direction. In particular, one can contemplate a perturbation in form
of a baby-Skyrmion string which can trigger a run-away instability.
The reason behind the instability is basically the requirement of the
radial direction to be special (that is, to obey only a second-order
equation of motion, whereas the angular directions enjoy many more
powers of derivatives). This loss of isotropy brings about the latter
mentioned instability. 

In this paper, we take the construction of higher-order derivative
corrections to the next level.
The spirit of our construction is similar to that behind the Skyrme
term and the BPS-Skyrme term.
Take the Skyrme term; it is fourth order in spacetime derivatives.
The most general term with fourth-order derivatives will contain four
\emph{time} derivatives. The Skyrme term does not; it is constructed
in such a way as to cancel the fourth-order derivatives in the $i$-th
space or time direction and contains four spacetime derivatives only
as a product of second-order derivatives in two different space or
time directions;
e.g. $(\p_x\phi)^2(\p_y\phi)^2$ or $(\p_t\phi)^2(\p_x\phi)^2$.
Note that this is the minimal number of derivatives in the $i$-th
direction (we will denote this number by $\delta$).
Two derivatives in the $i$-th direction is, however, only possible for
terms up to and including sixth-order in derivatives in 3 spatial
dimensions or eighth-order in derivatives in 4 spacetime
dimensions. We prove, however, that the latter term vanishes
identically in the Skyrme model ($S^3$ target space). 
For eighth-, tenth- and twelfth-order derivative terms, the smallest
number of derivatives in the $i$-th direction is four,
i.e.~$\delta=4$.
That is, when we do not break isotropy.

Our construction is straightforward and yields positive-definite
static energies for the systems. We find simple interpretations for
the Lagrangians that we constructed. The eighth-order Lagrangian can
be understood as the sum of the Skyrme-term squared and the kinetic
term multiplied by the BPS-Skyrme term (the sixth-order term mentioned
above). The tenth-order Lagrangian can be interpreted as the Skyrme
term multiplied by the BPS-Skyrme term.
Finally, the twelfth-order Lagrangian can be interpreted as the
BPS-Skyrme term squared. 

We successfully achieve manifest stability for static energy
associated with the higher-order Lagrangians.
However, in order to check that time-dependent perturbations cannot
spoil this stability, we construct the corresponding Hamiltonians.
The Hamiltonians, as well known, are important objects because they
give rise to the Euler-Lagrange equations of motion (as the
Lagrangians do) and because we do not have any explicit time
dependence, they are conserved and thus can be associated with the
total energy.
Although the Hamiltonians do not suffer from the famous Ostrogradsky 
instability \cite{Ostrogradsky:1850,Woodard:2015zca} (see also
Ref.~\cite{Pais:1950za}), their highly 
nonlinear nature induces nonlinearities in the conjugate momenta and
hence in the Hamiltonians themselves which potentially may destabilize
the systems and in turn their solitons.
The dynamical instability we find is intrinsically different from the
Ostrogradsky one, because we do not have two time derivatives acting
on the same field, but simply large powers of one time derivative
acting on one field (see Appendix \ref{app:diffOstrogradsky}). This
implies that we only have a single conjugate 
momentum for each field (as opposed to several as in Ostrogradsky's
Lagrangian) and there is no run-away associated with a
linear conjugate momentum in the Hamiltonian.
Nevertheless, our construction yields a nonlinear conjugate momentum
which induces ghost-like kinetic terms. In particular, the terms
containing fourth-order time derivatives are accompanied by two powers
of the inverse metric, which thus acquires the wrong sign -- this term
therefore remains negative in the Hamiltonian.
The other effect we find is also related to the nonlinearities of the
higher-order derivative terms, namely, when a term has more than two
time derivatives the SO(3,1) symmetry of the Lorentz invariants is not 
simply transformed to SO(4) invariants by the standard Legendre
transform, but the latter SO(4) symmetry is broken. This breaking of
the would-be SO(4) symmetry induces terms with both signs.
This is also related to our construction producing ``minimal''
Lagrangians, i.e.~terms that are as simple as possible in terms of
eigenvalues of the strain tensor. 
Although we find the above dynamical instabilities in our
Hamiltonians,
we conjecture that the vacuum is stable.

Finally, we argue that the Hamiltonian intrinsically knows that it is
a low-energy effective field theory and that the instabilities
described above do not occur at leading order for time-dependent
perturbations.
We consider the simplest possible perturbation, i.e.~exciting the
translational zero mode, and associating the energy scale of said
perturbation with a velocity.
We find exact conditions for when the instability sets in and estimate
the velocities for which the effective theory will break down.
In all cases the critical velocities are of the order of about half
the speed of light. 
Then we show that to leading order in
the velocity squared, there is no instability of the Hamiltonians of
eighth and twelfth order.

The paper is organized as follows.
In section \ref{sec:formalism}, we set up the formalism to construct
the higher-order derivative Lagrangians. 
In section \ref{sec:Marleau}, we review the Marleau construction and
show that it contains an instability already in the static energy. 
Section \ref{sec:positive_static_Ls} presents our construction of
higher-order derivative Lagrangians with positive-definite static
energy.
In section \ref{eq:Hamiltonians} the corresponding Hamiltonians are
then constructed and dynamical instabilities are found and
discussed. 
Section \ref{sec:low_energy_stability} then discusses the low-energy
stability of the Hamiltonians.
Section \ref{sec:discussion} then concludes with a discussion. 
Appendix \ref{app:runaway} illustrates the baby-Skyrmion string
triggering a run-away perturbation found in the Marleau construction
while Appendix \ref{app:diffOstrogradsky} provides a comparison of our
dynamical instability with that of Ostrogradsky and the differences in
their underlying Lagrangians.

\section{The formalism for higher-order terms}\label{sec:formalism}

Traditionally, the Skyrme model is formulated in terms of
left-invariant current $L_\mu\equiv U^\dag\p_\mu U$ (or equivalently the
right-invariant current $R_\mu\equiv\p_\mu U U^\dag$), $\mu=0,1,2,3$
is a spacetime index, where $U$ is the chiral Lagrangian field
\beq
U=\sigma\mathbf{1}_2+i\pi^a\tau^a \in \textrm{SU}(2),
\eeq
with $\tau^a$ the Pauli matrices, $a=1,2,3$, and $U$ obeys the
nonlinear sigma-model constraint $\det U=1$. 

The kinetic term is then simply given by
\beq
\mathcal{L}_2 = \frac{1}{4}\Tr(L_\mu L^\mu),
\eeq
and we are using the mostly-positive metric signature.
Both the Skyrme term, which is of fourth order in derivatives, and the
BPS-Skyrme term \cite{Adam:2010fg,Adam:2010ds}, which is of sixth
order in derivatives, is made out of antisymmetric combinations of
$L_\mu$, 
\begin{align}
\mathcal{L}_4 &= \frac{1}{32}\Tr[L_\mu,L_\nu][L^\mu,L^\nu]
  = -\frac{1}{32}\Tr[F_{\mu\nu} F^{\nu\mu}],
\label{eq:Skyrmeterm}\\
\mathcal{L}_6 &= \frac{1}{144}\eta_{\mu\mu'}
\epsilon^{\mu\nu\rho\sigma}\Tr[L_\nu L_\rho L_\sigma]
\epsilon^{\mu'\nu'\rho'\sigma'}\Tr[L_{\nu'} L_{\rho'} L_{\sigma'}]
  = \frac{1}{96}\Tr[F_{\mu}^{\phantom{\mu}\nu}
  F_{\nu}^{\phantom{\nu}\rho} F_{\rho}^{\phantom{\rho}\mu}],
\label{eq:BPSSkyrmeterm}
\end{align}
where we have defined
\beq
F_{\mu\nu} \equiv [L_\mu,L_\nu],
\eeq
and $\eta_{\mu\nu}$ is the flat-space Minkowski metric of
mostly-positive signature. 
Proving that the middle and right-hand side of
Eq.~\eqref{eq:BPSSkyrmeterm} are identical is somewhat nontrivial; we
will see that it is indeed the case after we switch to the notation of
eigenvalues, see below.

Although one can construct higher-order terms with more than six
derivatives using $F_{\mu\nu}$ (see Sec.~\ref{sec:Marleau}), it is
convenient to switch the notation to using invariants of O(4)
instead
\beq
\mathbf{n}_\mu\cdot\mathbf{n}_\nu
\equiv \p_\mu\mathbf{n}\cdot\p_\nu\mathbf{n},
\eeq
where
\beq
U = \mathbf{1}_2n^0 + in^a\tau^a,
\label{eq:nbasis}
\eeq
and the boldface symbol denotes the four vector
$\mathbf{n}\equiv(n^0,n^1,n^2,n^3)$ 
of unit length: $\mathbf{n}^2=1$.
This tensor is the strain tensor.

Since the Lagrangian is a Lorentz invariant, we can immediately see
that the simplest invariants of both O(4) and Lorentz symmetry we can  
write down, are given by
\beq
\ab{r}\equiv
\prod_{p=1}^r
\eta^{\mu_{p+1|r}\nu_p}\mathbf{n}_{\mu_p}\cdot\mathbf{n}_{\nu_p}
= (-2)^{-r} \prod_{p=1}^r
\eta^{\mu_{p+1|r}\nu_p}
\Tr[L_{\mu_p} L_{\nu_p}],
\label{eq:nninv}
\eeq
where the modulo function in the first index, $p+1|r$ (meaning $p+1$
mod $r$), simply ensures that the index $\mu_{r+1}$ is just $\mu_1$
and $\eta^{\mu\nu}$ is the inverse of the flat Minkowski metric of
mostly-positive signature. 

Another invariant of both SO(4) (which is a subgroup of O(4)) and of
Lorentz symmetry that we can construct is given by
\beq
\epsilon_{a b c d}\epsilon^{\mu\nu\rho\sigma}
  n_\mu^a n_\nu^b n_\rho^c n_\sigma^d,
\label{eq:epsilon_inv}
\eeq
which obviously vanishes for static fields.
Therefore, we can safely ignore this invariant for the static
solitons.

The most general static Lagrangian density with $2n$ derivatives, can
thus be written as 
\begin{align}
-\mathcal{L}_{2n} &=
\sum_{r_1=1,\ldots,n}\sum_{r_2=r_1,\ldots,n-r_1}\cdots
\sum_{r_n=r_{n-1},\ldots,n-\sum_{p=1,\ldots,(n-1)}r_p}
a_{r_n,r_{n-1},\ldots,r_1}\ab{r_n}\ab{r_{n-1}}\cdots\ab{r_1},
\label{eq:L2n_general}
\end{align}
where it is understood that a factor of $\ab{r_p}$ is only present when
the index $r_p$ has a positive range in the sum (including unity as
its only possibility).

The invariants \eqref{eq:nninv} with the hedgehog Ansatz
\beq
U = \mathbf{1}_2\cos f(\rho) + \frac{i x^a\tau^a}{\rho}\sin f(\rho),
\label{eq:hedgehog}
\eeq
have an astonishingly simple form
\beq
\ab{r} = f_\rho^{2r} + \frac{2\sin^{2r}f}{\rho^{2r}},
\eeq
where $f$ is a profile function with the boundary conditions
$f(\infty)=0$ and $f(0)=\pi$, $f_\rho\equiv \p_\rho f$ and
$\rho=\sqrt{(x^1)^2+(x^2)^2+(x^3)^2}$ is the radial coordinate.

It is, however, not enough to work with a spherically symmetric Ansatz
(i.e.~the hedgehog in Eq.~\eqref{eq:hedgehog}), as the system may have
runaway directions when not restricting to spherical symmetry.
It is clear that the static energy of the system is bounded from below
when all the coefficients $a\geq 0$ are positive semi-definite.
However, that case in general implies derivatives in one direction of
order $2n$.

In this paper, our philosophy will be similar to the construction of
the Skyrme term, namely we want to construct the higher-derivative
terms with the minimal number of derivatives in each spacetime
direction. 
That choice, however, implies that some of the coefficients $a$ need
to be negative. The prime example being the Skyrme term, for which we
have
\beq
a_{1,1} = -a_{2} = \frac{1}{4}.
\eeq
If we now consider $d$ spatial dimensions, the smallest possible
number of derivatives in the $i$-th direction (in the static case) is
given by 
\beq
\delta \equiv 2\lceil n/d\rceil,
\label{eq:deltadef}
\eeq
where $\lceil\chi\rceil=\mathop\textrm{ceil}(\chi)$ rounds the real
number $\chi$ up to its nearest integer.
This of course just corresponds to distributing the derivatives
symmetrically over all $d$ spatial dimensions. 
This means that for $d=3$, we can only have $\delta=2$ derivatives in
the $i$-th direction for $n\leq 3$, i.e.~at most six derivatives in
total. 
We can also see that if we consider $\delta=4$ derivatives in the
$i$-th direction, then $n=4,5,6$ yielding $8,10$, and $12$ derivative
terms. 
These are the terms we will focus on constructing in this paper. 

Since we now allow for some of the coefficients $a$ to be negative, we
have to find a method to ensure the stability of the system or in
other words positivity of the static energy of the system.
For this purpose, it will prove convenient to use the formalism of
eigenvalues \cite{Manton:1987xt} of the strain tensor
\beq
D_{ij} \equiv -\frac{1}{2} \Tr[L_i L_j]
= \mathbf{n}_i\cdot\mathbf{n}_j
= \left[V
\begin{pmatrix}
\lambda_1^2\\
&\lambda_2^2\\
&&\lambda_3^2
\end{pmatrix}
V^{\rm T}\right]_{ij},
\label{eq:lambdadef}
\eeq
which we will denote as
\beq
\lambda_1^2,\lambda_2^2,\lambda_3^2,
\eeq
$i,j=1,2,3$ and $V$ is an orthogonal matrix.
It is now easy to prove that
\beq
\ab{r} = \lambda_1^{2r} + \lambda_2^{2r} + \lambda_3^{2r}.
\label{eq:rlambda}
\eeq
This means that the invariant $\ab{r}$ has exactly the maximal number
(i.e.~$2r$) of derivatives in one direction (and due to symmetry this
term is summed over all spatial directions).

Now our construction works as follows.
We write down the most general Lagrangian density of order $2n$ using
Eq.~\eqref{eq:L2n_general}. 
Then we calculate the number of derivatives of $\mathbf{n}$ in one
direction, say $x^1$. The general case has $2n$ derivatives in the
$x^1$-direction. Finding the linear combinations with only $\delta$
(see Eq.~\eqref{eq:deltadef}) derivatives in the $x^1$-direction is
tantamount to solving the constraints of setting the coefficients of
the terms with $2n$, $2n-2$, $\cdots$, $\delta+2$ orders of
derivatives in the $x^1$-direction equal to zero.
The final step is to ensure that all terms provide positive
semi-definite static energy when written in terms of the eigenvalues
$\lambda_i$, see Eq.~\eqref{eq:lambdadef}. 
We will carry out the explicit calculation in
Sec.~\ref{sec:positive_static_Ls}.

\section{The Marleau construction}\label{sec:Marleau}

In this section we will review the construction of
Marleau \cite{Marleau:1989fh,Marleau:1990nh,Marleau:1991jk,Marleau:2000ia}
for higher-order derivative terms.
The $2n$-th order Lagrangians are given by\footnote{There is a
difference in a factor of two for these terms for $n>1$ as compared
to those of Eqs.~\eqref{eq:Skyrmeterm}
and \eqref{eq:BPSSkyrmeterm}. The latter normalization is conventional
while the normalization below is chosen such that Eq.~\eqref{eq:L2nf}
holds.}  
\begin{align}
\mathcal{L}_2 &= \frac{1}{4} \eta^{\mu\mu'} \Tr(L_\mu L_{\mu'}),
  \label{eq:L2}\\
\mathcal{L}_4 &= -\frac{1}{64} \eta^{\mu\mu'} \eta^{\nu\nu'}
  \Tr(F_{\mu\nu'}F_{\nu\mu'}),\\ 
\mathcal{L}_6 &= \frac{1}{192} \eta^{\mu\mu'} \eta^{\nu\nu'} \eta^{\rho\rho'}
  \Tr(F_{\mu\nu'}F_{\nu\rho'}F_{\rho\mu'}),\\
\mathcal{L}_8 &= -\frac{1}{512} \eta^{\mu\mu'} \eta^{\nu\nu'} \eta^{\rho\rho'}
  \eta^{\sigma\sigma'}\left[
  \Tr(F_{\mu\nu'}F_{\nu\rho'}F_{\rho\sigma'}F_{\sigma\mu'})
  - \Tr\left(\left\{F_{\mu\nu'},F_{\rho\sigma'}\right\}F_{\nu\rho'}F_{\sigma\mu'}\right)
  \right], \\
\mathcal{L}_{10} &= 0,\\
\mathcal{L}_{12} &= \frac{1}{6144} \eta^{\mu\mu'} \eta^{\nu\nu'} \eta^{\rho\rho'}
  \eta^{\sigma\sigma'} \eta^{\lambda\lambda'} \eta^{\delta\delta'}\bigg[
  \Tr(F_{\mu\nu'}F_{\nu\rho'}F_{\rho\sigma'}F_{\sigma\lambda'}F_{\lambda\delta'}F_{\delta\mu'})
  \label{eq:L12}\\
  &\phantom{=\ }
  -\frac{9}{2}\Tr\left(\left\{F_{\mu\nu'},F_{\rho\sigma'}\right\}F_{\nu\rho'}F_{\sigma\lambda'}F_{\lambda\delta'}F_{\delta\mu'}\right) 
  +\frac{7}{2}\Tr\left(\left\{F_{\mu\nu'},F_{\rho\sigma'}\right\}\left\{F_{\nu\rho'},F_{\lambda\delta'}\right\}F_{\sigma\lambda'}F_{\delta\mu'}\right)
  \bigg]. \nonumber
\end{align}
The first three Lagrangians already have at most two derivatives in
one direction $\delta=2$, as we have seen in the previous section.
Starting from the eight-order derivative term ($n=4$), the systematic
construction works like this. Take $n$ $F$-factors and contract their
Lorentz indices as a matrix product and then subtract the following
terms: the first one is made by switching the second and the third $F$
and then anti-commuting the first and the new second $F$ (the old
third $F$ at position 2). The next term starts with the previous term
and switches the fourth and fifth $F$ and then anti-commutes the third
and new fourth $F$ (the old fifth $F$ at position 4). This continues
as long as there are enough $F$ factors to keep on going.

Notice, that this construction cannot produce a tenth-order derivative
term as it vanishes identically. 

Although the Lagrangian densities (\ref{eq:L2}-\ref{eq:L12}) seem
overly complicated in terms of the Skyrme term, Marleau found that for
the hedgehog Ansatz, they simplify
drastically \cite{Marleau:1989fh,Marleau:1990nh,Marleau:1991jk,Marleau:2000ia} 
to
\beq
\mathcal{L}_{2n} = -\frac{\sin^{2n-2}(f)}{2\rho^{2n-2}} f_\rho^2
  -\frac{3-n}{2n}\frac{\sin^{2n}(f)}{\rho^{2n}}.
\label{eq:L2nf}
\eeq
Notice, however, that for $n>3$ the second term in this reduced
Lagrangian density is negative definite (since $\rho\geq 0$ and
$\sin f\geq 0$ are both positive semi-definite).

By explicit calculation, we find by plugging Eq.~\eqref{eq:nbasis}
into the Lagrangians (\ref{eq:L2}-\ref{eq:L12})
\begin{align}
\mathcal{L}_2 &= -\frac{1}{2}\ab{1},
  \label{eq:MarleauL2inv}\\
\mathcal{L}_4 &= \frac{1}{8}\ab{2} - \frac{1}{8}\ab{1}^2,\\
\mathcal{L}_6 &= -\frac{1}{6}\ab{3} + \frac{1}{4}\ab{2}\ab{1}
  - \frac{1}{12}\ab{1}^3,
  \label{eq:MarleauL6inv}\\
\mathcal{L}_8 &= \frac{13}{16}\ab{4} - \frac{5}{4}\ab{3}\ab{1}
  - \frac{3}{8}\ab{2}^2 + \ab{2}\ab{1}^2 - \frac{3}{16}\ab{1}^4,
  \label{eq:MarleauL8inv}\\
\mathcal{L}_{10} &= 0, \\
\mathcal{L}_{12} &= \frac{55}{24}\ab{6} - \frac{11}{2}\ab{5}\ab{1}
  - \frac{11}{8}\ab{4}\ab{2} + \frac{35}{8}\ab{4}\ab{1}^2
  - \frac{13}{24}\ab{3}^2
  + \frac{29}{8}\ab{3}\ab{2}\ab{1} - \frac{47}{24}\ab{3}\ab{1}^3 \non
&\phantom{=\ }
  + \frac{1}{12}\ab{2}^3 
  - \frac{3}{2}\ab{2}^2\ab{1}^2 + \frac{1}{2}\ab{2}\ab{1}^4,
  \label{eq:MarleauL12inv}
\end{align}
where we have used
\begin{align}
F_{\mu\nu} &= -2iX_{\mu\nu}^a\tau^a,\qquad&
X_{\mu\nu}^a &= \epsilon^{abc} n_\mu^b n_\nu^c + n_\mu^0 n_\nu^a -
n_\mu^a n_\nu^0,\\
L_\mu &= -iX_\mu^a \tau^a,\qquad&
X_\mu^a &= \epsilon^{abc} n_\mu^b n^c + n_\mu^0 n^a -
n_\mu^a n^0,
\end{align}
and the contraction
\beq
X_{\mu\nu}^a X_{\rho\sigma}^a =
(\mathbf{n}_\mu\cdot\mathbf{n}_\rho)(\mathbf{n}_\nu\cdot\mathbf{n}_\sigma)
-(\mathbf{n}_\mu\cdot\mathbf{n}_\sigma)(\mathbf{n}_\nu\cdot\mathbf{n}_\rho)
+\epsilon^{abcd} n_\mu^a n_\nu^b n_\rho^c n_\sigma^d.
\eeq
An easy check that one can make is to sum all the coefficients $a$ in
each Lagrangian density and see that indeed the sum vanishes for all 
$\mathcal{L}_{2n}$ with $n>1$. This simply means that the highest
power of derivatives vanishes for each of the higher-order Lagrangian
densities.

We can see from the reduced Lagrangian density \eqref{eq:L2nf}, that
for $n>3$, corresponding to 8 or more derivatives, the non-radial
derivative term (it is a combination of angular derivatives) acquires
a negative sign. Since $0\leq \sin f\leq 1$ for the profile function
$f$ in the range $f\in[0,\pi]$, there is no runaway asymptotically.
Nevertheless, a negative sign in the energy could signal some runaway
instabilities that are just not allowed for by the spherically
symmetric Ansatz \eqref{eq:hedgehog}.
In fact, for the hedgehog Ansatz, Ref.~\cite{Jackson:1991mt}
found a stability criterion for the Marleau construction.

In order to understand the instabilities in the Marleau construction, 
we take the Lagrangian densities written in terms of the invariants,
i.e.~Eqs.~(\ref{eq:MarleauL2inv}-\ref{eq:MarleauL12inv}) and plug in
the relation \eqref{eq:rlambda} 
\begin{align}
\mathcal{L}_2 &= -\frac{1}{2}\left(\lambda_1^2 + \lambda_2^2
  + \lambda_3^2\right),\\
\mathcal{L}_4 &= -\frac{1}{4}\left(\lambda_1^2\lambda_2^2
  + \lambda_1^2\lambda_3^2 + \lambda_2^2\lambda_3^2\right),\\
\mathcal{L}_6 &= -\frac{1}{2}\lambda_1^2\lambda_2^2\lambda_3^2,\\
\mathcal{L}_8 &= -\frac{1}{8}\left(
   2\lambda_1^4\lambda_2^2\lambda_3^2
  +2\lambda_1^2\lambda_2^4\lambda_3^2
  +2\lambda_1^2\lambda_2^2\lambda_3^4
  -\lambda_1^4\lambda_2^4
  -\lambda_1^4\lambda_3^4
  -\lambda_2^4\lambda_3^4\right),\\
\mathcal{L}_{10} &= 0,\\
\mathcal{L}_{12} &= -\frac{1}{4}\big(
  \lambda_1^6\lambda_2^4\lambda_3^2
  +\lambda_1^4\lambda_2^2\lambda_3^6
  +\lambda_1^2\lambda_2^6\lambda_3^4
  +\lambda_1^2\lambda_2^4\lambda_3^6
  +\lambda_1^4\lambda_2^6\lambda_3^2
  +\lambda_1^6\lambda_2^2\lambda_3^4 \non
&\phantom{=-\frac{1}{4}\big(\ }
  -2\lambda_1^4\lambda_2^4\lambda_3^4
  -\lambda_1^6\lambda_2^6
  -\lambda_1^6\lambda_3^6
  -\lambda_2^6\lambda_3^6\big).
\end{align}
Clearly the construction yields non-manifestly positive terms for the
eighth- and twelfth-order Lagrangians. 
We can see the trend that most terms that are products of derivatives
in all three spatial dimensions are positive, whereas all terms that
are products of derivatives in two spatial dimensions are
negative.\footnote{In Ref.~\cite{Longpre:2005fr,Longpre:2007zz} a
negative coefficient of the eighth-order Lagrangian was used to avoid
the baby-Skyrmion string instability; that unfortunately yields a
potential instability due to Derrick collapse of the entire soliton. }

It is easy to construct a perturbation that can drive the system into
a runaway direction. Consider a perturbation that depends only on
$x^1,x^2$ but not on $x^3$, then it is clear that for such
perturbation the static energies for different Lagrangian densities
become 
\begin{equation}
-\mathcal{L}_2 = \frac{1}{2}\left(\lambda_1^2+\lambda_2^2\right),\quad
-\mathcal{L}_4 = \frac{1}{4}\lambda_1^2\lambda_2^2,\quad
-\mathcal{L}_6 = 0,\quad
-\mathcal{L}_8 = -\frac{1}{8}\lambda_1^4\lambda_2^4,\quad
-\mathcal{L}_{12} = -\frac{1}{4}\lambda_1^6\lambda_2^6.
\label{eq:baby-Skyrme-instability}
\end{equation}
For illustrative purposes, we will show an example of a run-away in 
Appendix \ref{app:runaway}.

Let us contemplate for a moment what the Marleau construction does.
It is clear that the $\ab{r}$-invariants themselves have a symmetric
distribution of derivatives in all spatial directions. There are
therefore no preferred direction per se.
Nevertheless, the Marleau construction is able to eliminate all terms
with $f_\rho^{2p}$ for $p>1$ and therefore the other $2n-2$ derivatives
must necessarily be angular derivatives. Since there are only two
angular directions in 3 dimensional space, there must be more than two
derivatives in at least one of the angular directions when $n>3$.
The way it works is to take the Lagrangian with $2n$ derivatives,
$\mathcal{L}_{2n}$, say using Eq.~\eqref{eq:L2n_general} and expand it
in powers of $f_\rho^2$. Then set the combinations of the coefficients
$a$ to zero for all terms with higher powers of $f_\rho^2$.

One may ask whether the Marleau construction is unique and more
importantly whether there exists a construction for higher-derivative
terms with more than six derivatives, that can provide at most two
radial derivatives (i.e.~at most $f_\rho^2$) and in the same time a
positive-definite static energy. 
To answer this, let us count how many parameters are left free
by the constraints setting terms with $f_\rho^k=0$ for $k>2$. 
Table \ref{tab:numinvparm} lists the number of free parameters for a
Lagrangian density with $2n$ derivatives. We have used one parameter
to normalize the second-order radial derivative term.
\begin{table}[!ht]
\begin{center}
\begin{tabular}{rccc}
$(2n)$ & invariants & constraints & free parameters\\
\hline\hline
2  & 1  & 0 & 0\\
4  & 2  & 1 & 0\\
6  & 3  & 2 & 0\\
8  & 5  & 3 & 1\\
10 & 7  & 4 & 2\\
12 & 11 & 5 & 5
\end{tabular}
\caption{Number of derivatives, O(4) and Lorentz invariants,
constraints and free parameters in the Marleau construction. } 
\label{tab:numinvparm}
\end{center}
\end{table}
Note that the number of invariants is indeed the partition function of
$n$ (in number theory). 
Notice however that the free parameters merely allow one to write the
same Lagrangian using different combinations of invariants (this
should be straightforward from the point of view of group theory).
Once the overall normalization is fixed, there are no free parameters
left. In order to demonstrate this last point, let us construct the
Lagrangians $\mathcal{L}_{2n}$ for $n=4,5,6$ explicitly
\begin{align}
-\mathcal{L}_8 &=
a_4 \ab{4} + a_{3,1} \ab{3}\ab{1}
-\left(\frac{3}{4}a_4 + \frac{3}{16}a_{3,1}\right)\ab{2}^2
-\left(\frac{1}{2}a_4 + \frac{9}{8}a_{3,1}\right)\ab{2}\ab{1}^2 \non
&\phantom{=\ }
+\left(\frac{1}{4}a_4 + \frac{5}{16}a_{3,1}\right)\ab{1}^4,
\label{eq:L8M_general}\\
-\mathcal{L}_{10} &=
a_5 \ab{5} + a_{4,1} \ab{4}\ab{1} + a_{3,2} \ab{3}\ab{2}
+\left(\frac{5}{3}a_5 - \frac{4}{3}a_{4,1} +
3a_{3,2}\right)\ab{3}\ab{1}^2 \non
&\phantom{=\ }
-\left(\frac{15}{8}a_5 + \frac{1}{2}a_{4,1}
  + \frac{9}{4}a_{3,2}\right)\ab{2}^2\ab{1} 
-\left(\frac{5}{4}a_5 - a_{4,1}
  + \frac{5}{2}a_{3,2}\right)\ab{2}\ab{1}^3 \non
&\phantom{=\ }
+\left(\frac{11}{24}a_5 - \frac{1}{6}a_{4,1}
  + \frac{3}{4}a_{3,2}\right)\ab{1}^5,
\label{eq:L10M_general}
\end{align}
\begin{align}
-\mathcal{L}_{12} &=
a_6 \ab{6} + a_{5,1} \ab{5}\ab{1} + a_{4,2} \ab{4}\ab{2}
+ a_{4,1,1} \ab{4}\ab{1}^2
+ a_{3,3} \ab{3}^2 + a_{3,2,1} \ab{3}\ab{2}\ab{1} \non
&\phantom{=\ }
+\left(\frac{10}{9}a_6 - \frac{5}{27}a_{5,1} + \frac{28}{27}a_{4,2}
  - \frac{4}{3}a_{4,1,1} + 2a_{3,3}
  + \frac{7}{9}a_{3,2,1}\right)\ab{3}\ab{1}^3 \non
&\phantom{=\ }
-\left(\frac{23}{48}a_6 + \frac{25}{288}a_{5,1} + \frac{23}{36}a_{4,2}
  + \frac{3}{8}a_{3,3} + \frac{5}{48}a_{3,2,1}\right)\ab{2}^3 \non
&\phantom{=\ }
-\left(\frac{67}{48}a_6 + \frac{365}{288}a_{5,1}
  + \frac{37}{36}a_{4,2} + \frac{1}{2}a_{4,1,1} + \frac{15}{8}a_{3,3}
  + \frac{73}{48}a_{3,2,1}\right)\ab{2}^2\ab{1}^2 \non
&\phantom{=\ }
-\left(\frac{7}{16}a_6 - \frac{55}{96}a_{5,1} + \frac{7}{12}a_{4,2}
  -a_{4,1,1} + \frac{9}{8}a_{3,3}
  + \frac{5}{16}a_{3,2,1}\right)\ab{2}\ab{1}^4 \non
&\phantom{=\ }
+\left(\frac{29}{144}a_6 - \frac{29}{864}a_{5,1}
  + \frac{23}{108}a_{4,2} - \frac{1}{6}a_{4,1,1} + \frac{3}{8}a_{3,3}
  + \frac{23}{144}a_{3,2,1}\right)\ab{1}^6.
\label{eq:L12M_general}
\end{align}
These are the most general Lagrangians with 1,2 and 5 free parameters,
respectively, that give rise to the radial Lagrangian \eqref{eq:L2nf}
with the coefficients
\begin{equation}
c_8 = 8a_4 + 6a_{3,1}, \qquad
c_{10} = 25a_5 + 30a_{3,2}, \qquad
c_{12} = 30a_6 + 15a_{5,1} + 24a_{4,2} + 36a_{3,3} + 18a_{3,2,1},
\label{eq:c8c10c12}
\end{equation}
and with the characteristic of having only two radial derivatives (by 
construction of course).
The Lagrangians in Eqs.~\eqref{eq:MarleauL8inv}
and \eqref{eq:MarleauL12inv} correspond to setting $a_4=-13/16$,
$a_{3,1}=5/4$ and $a_6=-55/24$, $a_{5,1}=11/2$, $a_{4,2}=11/8$,
$a_{4,1,1}=-35/8$, $a_{3,3}=13/24$, $a_{3,2,1}=-29/8$, respectively. 
The simplest possible Lagrangians can be written by setting the
1, 2 and 5 coefficients of the largest invariants to zero
\begin{align}
\mathcal{L}_8 &= -a_{3,1} \left(
  \ab{3}\ab{1} - \frac{3}{16} \ab{2}^2 - \frac{9}{8} \ab{2}\ab{1}^2
  + \frac{5}{16}\ab{1}^4\right), \label{eq:Marleau8simpl}\\ 
\mathcal{L}_{10} &= -a_{3,2} \left(
  \ab{3}\ab{2} + 3 \ab{3}\ab{1}^2 - \frac{9}{4} \ab{2}^2\ab{1}
  - \frac{5}{2} \ab{2}\ab{1}^3
  + \frac{3}{4} \ab{1}^5\right),\\
\mathcal{L}_{12} &= -a_{3,2,1} \left(
  \ab{3}\ab{2}\ab{1}  + \frac{7}{9} \ab{3}\ab{1}^3
  - \frac{5}{48} \ab{2}^3
  - \frac{73}{48} \ab{2}^2\ab{1}^2 - \frac{5}{16} \ab{2}\ab{1}^4
  + \frac{23}{144} \ab{1}^6\right).
\end{align}
In order to normalize the above Lagrangian densities like
Eq.~\eqref{eq:L2nf}, we need to set $a_{3,1}=1/6$, $a_{3,2}=1/30$ and
$a_{3,2,1}=1/18$, respectively.

Note that the highest invariant we need to describe these higher-order
Lagrangians is the $\ab{3}$, which is the chain-contraction of the
Lorentz indices of three O(4) invariants.

In order to see whether the free parameters can change the Lagrangian
densities, we rewrite
Eqs.~(\ref{eq:L8M_general}-\ref{eq:L12M_general}) using the
relation \eqref{eq:rlambda}, obtaining
\begin{align}
\mathcal{L}_8 &= -\frac{c_8}{8}\left(
  2\lambda_1^4\lambda_2^2\lambda_3^2
  +2\lambda_1^2\lambda_2^4\lambda_3^2
  +2\lambda_1^2\lambda_2^2\lambda_3^4
  -\lambda_1^4\lambda_2^4
  - \lambda_1^4\lambda_3^4
  -\lambda_2^4\lambda_3^4\right), \\
\mathcal{L}_{10} &= -\frac{c_{10}}{10}\big(
  2\lambda_1^6\lambda_2^2\lambda_3^2
  +2\lambda_1^2\lambda_2^6\lambda_3^2
  +2\lambda_1^2\lambda_2^2\lambda_3^6
  +\lambda_1^4\lambda_2^4\lambda_3^2
  +\lambda_1^4\lambda_2^2\lambda_3^4
  +\lambda_1^2\lambda_2^4\lambda_3^4 \non
&\phantom{=-c_{10}\ \ }
  -\lambda_1^6(\lambda_2^4 + \lambda_3^4)
  -\lambda_2^6(\lambda_1^4 + \lambda_3^4)
  -\lambda_3^6(\lambda_1^4 + \lambda_2^4)
  \big),
\end{align}
\begin{align}
\mathcal{L}_{12} &=
-\left(\frac{11}{108}c_{12} - \frac{2}{3}\tilde{c}_{12}\right)\big( 
  \lambda_1^6(\lambda_2^4\lambda_3^2 + \lambda_2^2\lambda_3^4) 
  +\lambda_2^6(\lambda_1^4\lambda_3^2 + \lambda_1^2\lambda_3^4) 
  +\lambda_3^6(\lambda_1^4\lambda_2^2 + \lambda_1^2\lambda_2^4) 
  -\lambda_1^6\lambda_2^6 \non
&\phantom{=-(108c_{12}-2\tilde{c}_{12})\qquad }
  -\lambda_1^6\lambda_3^6
  -\lambda_2^6\lambda_3^6\big) \non
&\phantom{=\ }
-\left(\frac{2}{27}c_{12} + \frac{1}{3}\tilde{c}_{12}\right)\big( 
  2\lambda_1^8\lambda_2^2\lambda_3^2
  +2\lambda_1^2\lambda_2^8\lambda_3^2
  +2\lambda_1^2\lambda_2^2\lambda_3^8 
  -\lambda_1^8(\lambda_2^4 + \lambda_3^4)
  -\lambda_2^8(\lambda_1^4 + \lambda_3^4) \non
&\phantom{=-(28c_{12}+3\tilde{c}_{12})\big(\quad \ \ }
  -\lambda_3^8(\lambda_1^4 + \lambda_2^4)\big) \non
&\phantom{=\ }
  +\left(\frac{1}{18}c_{12} - 2\tilde{c}_{12}\right)
  \lambda_1^4\lambda_2^4\lambda_3^4,
\label{eq:L12delta2general}
\end{align}
where the coefficients $c_{8,10,12}$ are given in
Eq.~\eqref{eq:c8c10c12} and we have defined
\beq
\tilde{c}_{12} \equiv a_{3,3} + \frac{1}{3}a_6.
\eeq
Notice that the eighth-order and tenth-order Lagrangians,
$\mathcal{L}_{8,10}$ depend on the combinations given in
Eq.~\eqref{eq:c8c10c12}, which is just an overall normalization
coefficient.
The twelfth-order Lagrangian, on the other hand, has a residual free
parameter, $\tilde{c}_{12}$. Say if we fix $c_{12}$ of
Eq.~\eqref{eq:c8c10c12} to one, then we still have a one-parameter
family of Lagrangians with different eigenvalues $\lambda_i$ all
giving rise to the reduced radial Lagrangian \eqref{eq:L2nf} upon
using the hedgehog Ansatz \eqref{eq:hedgehog}.

As for stability, it is clear that for $\mathcal{L}_{8,10}$ all the
free parameters just give rise to the same Lagrangian with
normalization $c_{8,10}$ and hence the negative terms cannot be
eliminated.
For the twelfth-order Lagrangian, we have two parameters and two terms
(the two first terms in Eq.~\eqref{eq:L12delta2general}) that contain
negative terms. However, eliminating both the first and the second
term in the Lagrangian also kills the last term. Therefore for these
three Lagrangian densities, there is no way of constructing stable
static eighth-, tenth-, and twelfth-order Lagrangians with only
second-order radial derivatives for the hedgehog
Ansatz \eqref{eq:hedgehog}.
By stable we mean that the static energy is bounded from below and
hence is stable against non-baryonic perturbations, i.e.~perturbations
with vanishing baryon number.

\section{Positive-definite static energy for minimal
Lagrangians}\label{sec:positive_static_Ls} 

In this section we will require positive-definite static energy and
construct terms with eight and more derivatives. As shown in
Eq.~\eqref{eq:deltadef}, the smallest possible number of derivatives
in the $i$-th direction is 4 for $n=4,5,6$, corresponding to the
eighth-, tenth-, and twelfth-order Lagrangians.

\subsection{2, 4 and 6 derivatives}
As a warm-up, let us rederive the kinetic, the Skyrme term and the
BPS-Skyrme term.
The difference for these terms with respect to the higher-order terms
with 8, 10 and 12 derivatives, is that $\delta=2$ for the second-,
fourth- and sixth-order derivative term. This means that we can
consistently have only 2 derivatives in the $i$-th direction (whereas
for 8-12 derivatives, we need $\delta=4$). 

First, the kinetic term is trivial as it has only one possibility,
i.e., 
\beq
-\mathcal{L}_2 = a_1 \ab{1}
= a_1 (\lambda_1^2 + \lambda_2^2 + \lambda_3^2).
\eeq

Second, the Skyrme term is the simplest and first nontrivial
example. We start by writing
\beq
-\mathcal{L}_4 = a_2 \ab{2} + a_{1,1} \ab{1}^2.
\eeq
To eliminate the fourth-order derivatives in the $i$-th direction, we
set $-a_2=a_{1,1}=\frac{1}{2}c_{4|2,2}$ and arrive at
\beq
-\mathcal{L}_4 = \frac{c_{4|2,2}}{2} (-\ab{2} + \ab{1}^2)
= c_{4|2,2} (\llij1222 + \llij1322 + \llij2322).
\label{eq:rSkyrmeterm}
\eeq

Finally, let us rederive the BPS-Skyrme term.
The most general form is
\beq
-\mathcal{L}_6 = a_3 \ab{3} + a_{2,1} \ab{2}\ab{1} + a_{1,1,1} \ab{1}^3.
\eeq
Eliminating the sixth-order derivatives in the $i$-th direction yields
the constraint
\beq
a_3 + a_{2,1} + a_{1,1,1} = 0,
\eeq
while eliminating the fourth-order yields
\beq
a_{2,1} + 3a_{1,1,1} = 0.
\eeq
Their common solution is simply
$a_{2,1}=-\frac{3}{2}a_3=-\frac{1}{2}c_{6|2,2,2}$ and
$a_{1,1,1}=\frac{1}{2}a_3=\frac{1}{6}c_{6|2,2,2}$.
Thus we obtain
\beq
-\mathcal{L}_6 = \frac{c_{6|2,2,2}}{3}
\left(\ab{3} - \frac{3}{2}\ab{2}\ab{1} + \frac{1}{2}\ab{1}^3\right)
= c_{6|2,2,2} \llls222.
\label{eq:rBPSSkyrmeterm}
\eeq
We are now ready to move on to the more complicated higher-order
derivative terms.

\subsection{8 derivatives}

Let us start with constructing the eighth-order Lagrangian. The most
general static Lagrangian can be written down as
\beq
-\mathcal{L}_8 =
  a_{4}\ab{4}
+ a_{3,1}\ab{3}\ab{1}
+ a_{2,2}\ab{2}^2
+ a_{2,1,1}\ab{2}\ab{1}^2
+ a_{1,1,1,1}\ab{1}^4.
\eeq
This Lagrangian density, however contains generally eight derivatives
in the same direction; therefore, we will constrain the Lagrangian
such that it has the minimal number of derivatives in each direction;
that is after constraining the above Lagrangian it will only contain
terms with at most 4 derivatives in the $i$-th direction.

Note that the above Lagrangian is constructed exactly as a sum over
all possible Ferrers diagrams in number theory or equivalently as a
sum over all possible Young tableaux with the total number of boxes
equal to $n$ (i.e.~four here) \cite{Andrews:1998}. Each row in the
Young tableau is identified with the O(4) invariant.

We will hence, eliminate all terms with 8 and 6 derivatives in the
(same) $i$-th direction; that is we allow for terms such as
$\lambda_1^4\lambda_2^4$, but will eliminate terms such as
$\lambda_1^8$ or $\lambda_1^6\lambda_2^2$.
We choose to solve these two constraints by eliminating $a_{1,1,1,1}$
and $a_{2,1,1}$, arriving at
\begin{align}
-\mathcal{L}_8
&= c_{8|4,4}
\left(\lambda_1^4 \lambda_2^4
+ \lambda_2^4 \lambda_3^4
+ \lambda_1^4 \lambda_3^4\right) 
+ c_{8|4,2,2}
\left(\lambda_1^4 \lambda_2^2 \lambda_3^2
+ \lambda_1^2 \lambda_2^4 \lambda_3^2
+ \lambda_1^2 \lambda_2^2 \lambda_3^4\right),
\label{eq:L8lambdadelta4general}
\end{align}
where we have defined
\beq
c_{8|4,4} \equiv 2a_4 + 4a_{2,2}, \qquad
c_{8|4,2,2} \equiv 8a_4 + 3a_{3,1} + 8a_{2,2}.
\label{eq:c8gen_coeff}
\eeq
Thus the 3 free parameters $a_4$, $a_{3,1}$ and $a_{2,2}$ only appear
in the above two combinations. 
Finally, in order to ensure stability of the static solutions, we
require that both $c_{8|4,4}>0$ and $c_{8|4,2,2}>0$.

Physically, we can interpret the two terms as follows. The first term
is the quadratic parts of the Skyrme term squared and the second term 
is the cross terms; in particular the whole Lagrangian is the Skyrme
term squared for $c_{8|4,2,2}=2c_{8|4,4}$. 
The second term also has a different interpretation than being the
cross terms from the squared Skyrme term; it is simply the
BPS-Skyrme term $\mathcal{L}_6$ multiplied by the Dirichlet term 
$\mathcal{L}_2$. Since the BPS-Skyrme term is the baryon charge
density squared, the latter term vanishes wherever the baryon charge
does. 

Writing the above Lagrangian in terms of the O(4) invariants, we get
\begin{align}
-\mathcal{L}_8 &=
a_4 \ab{4} + a_{3,1} \ab{3}\ab{1} + a_{2,2} \ab{2}^2
- \left(2a_4 + \frac{3}{2}a_{3,1} + 2a_{2,2}\right) \ab{2}\ab{1}^2
\non&\phantom{=\ }
+ \left(a_4 + \frac{1}{2}a_{3,1} + a_{2,2}\right) \ab{1}^4.
\label{eq:L8rdelta4general}
\end{align}
As an example, we can set $a_4=a_{3,1}=0$ to get the following minimal
Lagrangian
\beq
-\mathcal{L}_8^{\rm min} =
a_{2,2} \left(\ab{2} - \ab{1}^2\right)^2,
\eeq
which yields a manifestly positive static energy for $a_{2,2}>0$. 
This minimal Lagrangian is of course nothing but the Skyrme term
(Eq.~\eqref{eq:rSkyrmeterm}) squared.
As another example, we set $a_4=a_{2,2}=0$ obtaining
\beq
-\mathcal{L}_8^{2\times 6} =
a_{3,1} \ab{1}\left(\ab{3} - \frac{3}{2}\ab{2}\ab{1}
+ \frac{1}{2}\ab{1}^3\right),
\label{eq:L82times6}
\eeq
which is clearly the Dirichlet term multiplied by the BPS-Skyrme term,
see Eq.~\eqref{eq:rBPSSkyrmeterm}.
This was already clear from writing the Lagrangian in terms of the
eigenvalues in Eq.~\eqref{eq:L8lambdadelta4general}. We see, however,
also that if instead of setting $a_4=a_{2,2}=0$, we set
$a_{2,2}=-a_4/2$ then we get a nontrivial one-parameter family of
Lagrangians all described by the Dirichlet term multiplied by the
BPS-Skyrme term. This is not clear at all from the invariants and this
implies very nontrivial relations among the invariants.
For instance, we can write the exact same Lagrangian as
Eq.~\eqref{eq:L82times6} as
\beq
-\mathcal{L}_8^{2\times 6} =
a_4\left(\ab{4} - \frac{1}{2}\ab{2}^2 - \ab{2}\ab{1}^2
+ \frac{1}{2}\ab{1}^4\right),
\label{eq:L82times6again}
\eeq
which is the same if we normalize $a_4=\frac{3}{4}a_{3,1}$.

Plugging the hedgehog Ansatz \eqref{eq:hedgehog} into
Eq.~\eqref{eq:L8rdelta4general}, we get
\beq
-\mathcal{L}_8 =
c_{8|4,4} \frac{\sin^8(f)}{\rho^8}
+2c_{8|4,2,2} \frac{\sin^6(f)}{\rho^6} f_\rho^2
+(c_{8|4,2,2} + 2c_{8|4,4}) \frac{\sin^4(f)}{\rho^4} f_\rho^4,
\eeq
where the positive-definite coefficients are given in
Eq.~\eqref{eq:c8gen_coeff}. 
We can again see the physical interpretation that the terms with
coefficient $c_{8|4,2,2}$ are the BPS-Skyrme term multiplied by the 
kinetic (Dirichlet) term, while the two terms with coefficient
$c_{8|4,4}$ are the two terms of the Skyrme term squared individually. 
We can also clearly see how the Marleau construction manages to cancel
the third term in the above Lagrangian; setting
$c_{8|4,4}=-c_{8|4,2,2}/2$ accomplishes that at the expense of losing
the property of positive-definiteness of the static energy.

A final comment is in order. The eighth-order Lagrangian is the first 
Lagrangian which necessitates 4 powers of derivatives in the $i$-th
direction, that is, $\delta=4$. But it is also the first Lagrangian
that has two physically independent terms, as shown in
Eq.~\eqref{eq:L8lambdadelta4general}.

\subsection{10 derivatives}

We will now continue with the tenth-order Lagrangian. 
The most general static Lagrangian with 10 derivatives can be written
as 
\begin{align}
-\mathcal{L}_{10}
&= a_5 \ab{5}
+ a_{4,1} \ab{4}\ab{1}
+ a_{3,2} \ab{3}\ab{2}
+ a_{3,1,1} \ab{3}\ab{1}^2
+ a_{2,2,1} \ab{2}^2\ab{1}
+ a_{2,1,1,1} \ab{2}\ab{1}^3
\non&\phantom{=\ }
+ a_{1,1,1,1,1} \ab{1}^5.
\end{align}
Using Eq.~\eqref{eq:deltadef}, we find that again in this case, we
cannot have less than 4 derivatives in the $i$-th direction. 
We will eliminate all terms with 10, 8 and 6 derivatives in the (same)
$i$-th direction. Choosing to eliminate $a_{1,1,1,1,1}$,
$a_{2,1,1,1}$, $a_{2,2,1}$, and $a_{3,1,1}$, we get
\beq
-\mathcal{L}_{10} = c_{10|4,4,2}\left(
  \lambda_1^4\lambda_2^4\lambda_3^2
  +\lambda_1^4\lambda_2^2\lambda_3^4
  +\lambda_1^2\lambda_2^4\lambda_3^4
  \right),
\eeq
where we have defined
\beq
c_{10|4,4,2} \equiv -5a_5 - 6a_{3,2}.
\label{eq:c10gen_coeff}
\eeq
Thus the 2 free parameters only appear in one combination which is
fixed uniquely by normalization. 
Finally, as usual in this construction, we require $c_{10|4,4,2}>0$ to
be positive definite in order to ensure stability of the static
solutions. 

Notice that this tenth-order Lagrangian has only one term in
contradistinction to the eighth-order Lagrangian that is composed of
two physically distinct terms (in this construction of course).

Physically, there is a simple interpretation of the above
Lagrangian. It is simply the Skyrme term multiplied by the BPS-Skyrme
term. Since the BPS-Skyrme term is the baryon charge density squared,
this Lagrangian vanishes wherever the baryon charge does.

Writing the above Lagrangian in terms of the O(4) invariants, we
obtain
\begin{align}
-\mathcal{L}_{10} &= a_5 \ab{5} + a_{4,1} \ab{4}\ab{1}
+ a_{3,2} \ab{3}\ab{2}
-\left(\frac{5}{3}a_5 + \frac{4}{3}a_{4,1} + a_{3,2}\right)
  \ab{3}\ab{1}^2 \non
&\phantom{=\ }
-\left(\frac{5}{4}a_5 + \frac{1}{2}a_{4,1} + \frac{3}{2}a_{3,2}\right)
  \ab{2}^2\ab{1}
+\left(\frac{5}{2}a_5 + a_{4,1} + 2a_{3,2}\right) \ab{2}\ab{1}^3 \non
&\phantom{=\ }
-\left(\frac{7}{12}a_5 + \frac{1}{6}a_{4,1}
  + \frac{1}{2}a_{3,2}\right) \ab{1}^5.
\label{eq:L10geninv_nonsimpl}
\end{align}
Notice that although the coefficient $a_{4,1}$ appears in the above
formulation of the Lagrangian, it does not influence the normalization
coefficient $c_{10|4,4,2}$ given in Eq.~\eqref{eq:c10gen_coeff}. 
This is because when we write the invariants with the coefficient
$a_{4,1}$,
\beq
\ab{1}\left(\ab{4} - \frac{4}{3}\ab{3}\ab{1} - \frac{1}{2}\ab{2}^2
+ \ab{2}\ab{1}^2 - \frac{1}{6}\ab{1}^4\right) = 0,
\label{eq:nontrivial4rel}
\eeq
in terms of the eigenvalues, $\lambda_i$, we find that the above
expression vanishes identically.
This nontrivial relation among the invariants is in fact due to the
observation we made in the previous subsection for the eighth-order
Lagrangian, namely that the Dirichlet term multiplied by the
BPS-Skyrme term can be written in two apparently different ways:
Eq.~\eqref{eq:L82times6} and Eq.~\eqref{eq:L82times6again}.
Thus the above relation can simply be written as
\beq
\left(\ab{4} - \frac{1}{2}\ab{2}^2 - \ab{2}\ab{1}^2
+ \frac{1}{2}\ab{1}^4\right)
- \frac{4}{3}\ab{1}\left(\ab{3} - \frac{3}{2}\ab{2}\ab{1}
  + \frac{1}{2}\ab{1}^3\right) = 0,
\eeq
where the two terms are equal as we found in the previous subsection
and hence the nontrivial relation \eqref{eq:nontrivial4rel} follows.

Hence, we can simplify the Lagrangian to
\begin{align}
-\mathcal{L}_{10} &= a_5 \ab{5} + a_{3,2} \ab{3}\ab{2}
-\left(\frac{5}{3}a_5 + a_{3,2}\right) \ab{3}\ab{1}^2
-\left(\frac{5}{4}a_5 + \frac{3}{2}a_{3,2}\right) \ab{2}^2\ab{1} \non
&\phantom{=\ }
+\left(\frac{5}{2}a_5 + 2a_{3,2}\right) \ab{2}\ab{1}^3 
-\left(\frac{7}{12}a_5 + \frac{1}{2}a_{3,2}\right) \ab{1}^5.
\label{eq:L10geninv}
\end{align}
As an example, we can set $a_5=0$ and write the above Lagrangian as
\beq
-\mathcal{L}_{10}^{4\times 6} = -a_{3,2} \left(-\ab{2} + \ab{1}^2\right)
  \left(\ab{3} - \frac{3}{2}\ab{2}\ab{1} + \frac{1}{2}\ab{1}^3\right),
\label{eq:L10a32}
\eeq
from which it is clear that this is simply the Skyrme term
(Eq.~\eqref{eq:rSkyrmeterm}) multiplied by the BPS-Skyrme term
(Eq.~\eqref{eq:rBPSSkyrmeterm}).
The static energy is positive definite because $a_{3,2}<0$, see
Eq.~\eqref{eq:c10gen_coeff}. 
Since the entire Lagrangian \eqref{eq:c10gen_coeff} is simply the
Skyrme term multiplied by the BPS-Skyrme term, the complementary part
of the Lagrangian \eqref{eq:L10geninv} is also nontrivially equal to
the above expression. We can see how the other part looks like by
setting $a_{3,2}=0$, getting
\beq
-\mathcal{L}_{10}^{4\times 6} = a_5\left(
  \ab{5} - \frac{5}{3} \ab{3}\ab{1}^2 - \frac{5}{4} \ab{2}^2\ab{1}
  + \frac{5}{2} \ab{2}\ab{1}^3 - \frac{7}{12} \ab{1}^5
  \right).
\label{eq:L10a5}
\eeq
Using the eigenvalues, $\lambda_i$, we find that this Lagrangian is
equal to that of Eq.~\eqref{eq:L10a32} for
$a_5=\frac{6}{5}a_{3,2}<0$. 
This is again a highly nontrivial relation between different O(4)
invariants. 

Plugging the hedgehog Ansatz \eqref{eq:hedgehog} into
Eq.~\eqref{eq:L10geninv_nonsimpl}, we obtain
\beq
-\mathcal{L}_{10} = 2c_{10|4,4,2}
  \left(
  \frac{\sin^2(f)}{\rho^2} f_\rho^2
  +\frac{\sin^4(f)}{2\rho^4} 
  \right)
  \frac{\sin^4(f)}{\rho^4} f_\rho^2,
\eeq
where the positive-definite coefficient $c_{10|4,4,2}$ is given in
Eq.~\eqref{eq:c10gen_coeff}.
The physical interpretation is again very clear as the Skyrme term
multiplied by the BPS-Skyrme term.
It is also clear from the above construction why the tenth-order term
vanishes in the Marleau construction, because there is only one
coefficient and there is no way to eliminate the $f_\rho^4$ term
without setting the whole term to zero.

\subsection{12 derivatives}

The highest order in derivatives we will consider in this paper is
twelve.
The most general static Lagrangian density with 12 derivatives can be
written as
\begin{align}
-\mathcal{L}_{12}
&= a_6 \ab{6}
+ a_{5,1} \ab{5}\ab{1}
+ a_{4,2} \ab{4}\ab{2}
+ a_{4,1,1} \ab{4}\ab{1}^2
+ a_{3,3} \ab{3}^2
+ a_{3,2,1} \ab{3}\ab{2}\ab{1} \non
&\phantom{=\ }
+ a_{3,1,1,1} \ab{3}\ab{1}^3 
+ a_{2,2,2} \ab{2}^3
+ a_{2,2,1,1} \ab{2}^2\ab{1}^2
+ a_{2,1,1,1,1} \ab{2}\ab{1}^4
+ a_{1,1,1,1,1,1} \ab{1}^6.
\end{align}
Using Eq.~\eqref{eq:deltadef}, we find that this is the largest number of
derivatives in a term which cannot have less than $\delta=4$
derivatives in the $i$-th direction.  
Continuing along the lines of the previous subsections we eliminate
all terms with 12, 10, 8 and 6 derivatives in the (same) $i$-th
direction. Choosing to eliminate the coefficients $a_{1,1,1,1,1,1}$,
$a_{2,1,1,1,1}$, $a_{2,2,1,1}$, $a_{3,1,1,1}$, and $a_{2,2,2}$, we
arrive at
\beq
-\mathcal{L}_{12} = c_{12|4,4,4} \lambda_1^4\lambda_2^4\lambda_3^4,
\eeq
where we have defined
\beq
c_{12|4,4,4} \equiv 3a_6 + 9a_{3,3}.
\label{eq:c12gen_coeff}
\eeq
Thus the 2 free parameters only appear in the above combination which
is fixed once the normalization of this Lagrangian is.
As always in this construction, we require $c_{12|4,4,4}>0$ to be
positive definite in order to ensure stability of the static
solutions.

Physically, the interpretation of this Lagrangian is straightforward;
it is simply the BPS-Skyrme term squared or equivalently the
baryon-charge density to the fourth power. 

Writing this Lagrangian in terms of the O(4) invariants, we get
\begin{align}
-\mathcal{L}_{12} &= a_6 \ab{6} + a_{5,1} \ab{5}\ab{1}
  + a_{4,2} \ab{4}\ab{2} + a_{4,1,1} \ab{4}\ab{1}^2
  + a_{3,3} \ab{3}^2 \non
&\phantom{=\ }
-\left(2a_6 + \frac{5}{6}a_{5,1} + \frac{4}{3}a_{4,2} +
  3a_{3,3}\right) \ab{3}\ab{2}\ab{1}
-\left(\frac{5}{6}a_{5,1} + \frac{4}{3}a_{4,1,1} -
  a_{3,3}\right)\ab{3}\ab{1}^3 \non
&\phantom{=\ }
-\left(\frac{1}{4}a_6 + \frac{1}{2}a_{4,2}\right) \ab{2}^3
+\left(\frac{3}{2}a_6 + a_{4,2} - \frac{1}{2}a_{4,1,1}
  + \frac{9}{4}a_{3,3}\right) \ab{2}^2\ab{1}^2 \non
&\phantom{=\ }
-\left(\frac{1}{4}a_6 - \frac{5}{6}a_{5,1} + \frac{1}{6}a_{4,2} -
  a_{4,1,1} + \frac{3}{2}a_{3,3}\right) \ab{2}\ab{1}^4 \non
&\phantom{=\ }
-\left(\frac{1}{6}a_{5,1} + \frac{1}{6}a_{4,1,1}
  - \frac{1}{4}a_{3,3}\right) \ab{1}^6.
\label{eq:L12geninv_nonsimpl}
\end{align}
Notice that when the Lagrangian is written in terms of the
eigenvalues, $\lambda_i$, the only parameter is the overall
coefficient $c_{12|4,4,4}$ which is the
combination \eqref{eq:c12gen_coeff} of $a_6$ and $a_{3,3}$.
The 3 other parameters in the above Lagrangian thus have no influence
on the physics and so we again expect nontrivial relations among the
invariants.
They are
\beq
\ab{5} - \frac{5}{6} \ab{3}\ab{2} - \frac{5}{6} \ab{3}\ab{1}^2
  + \frac{5}{6} \ab{2}\ab{1}^3 - \frac{1}{6} \ab{1}^5 = 0,
\label{eq:nontrivial5rel}
\eeq
and Eq.~\eqref{eq:nontrivial4rel}, where the first is the relation
with coefficient $a_{5,1}$ and the latter appears with both
coefficients $a_{4,2}$ and $a_{4,1,1}$ as well as a factor of $\ab{2}$
and $\ab{1}^2$, respectively. The latter relation was discussed already
in the last subsection. 
The nontrivial relation Eq.~\eqref{eq:nontrivial5rel} can be
understood by writing it as
\begin{align}
\left(
\ab{5} - \frac{5}{3} \ab{3}\ab{1}^2 - \frac{5}{4} \ab{2}^2\ab{1}
  + \frac{5}{2} \ab{2}\ab{1}^3 - \frac{7}{12} \ab{1}^5
\right) \qquad
\non
\mathop- \frac{5}{6}\left(\ab{2}-\ab{1}^2\right)\left(
  \ab{3} - \frac{3}{2} \ab{2}\ab{1} + \frac{1}{2} \ab{1}^3
\right) = 0,
\end{align}
which is exactly the two Lagrangians \eqref{eq:L10a32}
and \eqref{eq:L10a5} with $a_{3,2}=\frac{5}{6}a_5$ and the nontrivial
relation \eqref{eq:nontrivial5rel} follows.

Hence, we can simplify the Lagrangian to
\begin{align}
-\mathcal{L}_{12} &= a_6 \ab{6} + a_{3,3} \ab{3}^2 
-\left(2a_6 + 3a_{3,3}\right) \ab{3}\ab{2}\ab{1}
+a_{3,3} \ab{3}\ab{1}^3 
-\frac{1}{4}a_6 \ab{2}^3 \non
&\phantom{=\ }
+\left(\frac{3}{2}a_6 + \frac{9}{4}a_{3,3}\right) \ab{2}^2\ab{1}^2 
-\left(\frac{1}{4}a_6 + \frac{3}{2}a_{3,3}\right) \ab{2}\ab{1}^4
+\frac{1}{4}a_{3,3} \ab{1}^6.
\label{eq:L12geninv}
\end{align}
As an example, we can set $a_6=0$ for which the above Lagrangian reads
\beq
-\mathcal{L}_{12}^{6\times 6} = a_{3,3}
\left(\ab{3} - \frac{3}{2} \ab{2}\ab{1} + \frac{1}{2} \ab{1}^3\right)^2,
\label{eq:L12a33}
\eeq
which is clearly the BPS-Skyrme term (Eq.~\eqref{eq:rBPSSkyrmeterm})
squared. 
Since the whole Lagrangian \eqref{eq:L12geninv} is the BPS-Skyrme
term squared, the complementary part -- i.e.~the part with coefficient
$a_6$ -- is also nontrivially the BPS-Skyrme term squared.
We can write that part down by setting $a_{3,3}=0$ in
Eq.~\eqref{eq:L12geninv}, yielding
\beq
-\mathcal{L}_{12}^{6\times 6} = a_6 \left(
  \ab{6} - 2 \ab{3}\ab{2}\ab{1} - \frac{1}{4} \ab{2}^3
  + \frac{3}{2} \ab{2}^2\ab{1}^2
  - \frac{1}{4} \ab{2}\ab{1}^4\right).
\eeq
Using the eigenvalues, $\lambda_i$, we find that the above Lagrangian
is exactly equal to that of Eq.~\eqref{eq:L12a33} for
$a_6=3a_{3,3}$. This is the last nontrivial relation we find between
different O(4) invariants. We expect the relation between these two
formulations of the twelfth-order Lagrangian to play a role in the
simplification of the fourteenth-order Lagrangian. 

Plugging the hedgehog Ansatz \eqref{eq:hedgehog} into the
Lagrangian \eqref{eq:L12geninv_nonsimpl}, we get
\beq
-\mathcal{L}_{12} = c_{12|4,4,4} \frac{\sin^8(f)}{\rho^8} f_\rho^4,
\eeq
where the positive-definite coefficient $c_{12|4,4,4}$ is given in
Eq.~\eqref{eq:c12gen_coeff}.
Again the physical interpretation is very clear as the above
expression is simply the baryon-charge density to the fourth power or
equivalently the BPS-Skyrme term squared.
In our construction, there is no term which is second order in
$f_\rho$ and so there is no overlap here between this construction and
the Marleau construction at this order.
It is clear why; in order to get a twelfth-order term with only two
radial derivatives, we need either 6 derivatives in the $\theta$
direction and 4 derivatives in the $\phi$ direction or vice versa.
Our construction eliminates such terms with 6 derivatives in the
$i$-the direction and hence, this term is not present in our
construction.

\section{Hamiltonians for the minimal Lagrangians}\label{eq:Hamiltonians}

In the last section we have constructed higher-order Lagrangians with
positive-definite static energies. This together with the nontrivial 
topological charge
\beq
\pi_3\left(\frac{{\rm SU}(2)\times{\rm SU}(2)}{{\rm SU}(2)}\right)
=\mathbb{Z},
\eeq
guarantees time-independent stability of the Skyrmions (solitons).
In this section, we will check that time-dependent perturbations are
also under control.
For this investigation, we need to calculate the Hamiltonians
corresponding to the Lagrangians obtained in the last section.

\subsection{Setup}

The first step is to compose the O(4) and Lorentz invariants into time
and spatial derivative parts, respectively. 
We thus define
\begin{align}
\langle r,0\rangle &\equiv \prod_{p=1}^r
  \mathbf{n}_{i_p}\cdot\mathbf{n}_{i_{p+1|r}}, \non
\langle r,1\rangle &\equiv (\mathbf{n}_0\cdot\mathbf{n}_{i_1})
  (\mathbf{n}_0\cdot\mathbf{n}_{i_{r-1}})
  \prod_{p=1}^{r-2} \mathbf{n}_{i_p}\cdot\mathbf{n}_{i_{p+1}}.
\label{eq:angular_brackets}
\end{align}
The first index $r$ in the brackets represents the number of spatial
indices in the product of invariants while the second represents the
number of time indices ($\mu=0$).
Notice that in the above angular brackets all indices are lowered.
Note also that there is no need for more than one time index in the
invariant, because two time indices always break the chain into two. 
Thus we can write the relevant Lorentz invariants as
\begin{align}
\ab{1} &= -\langle0,1\rangle +\langle1,0\rangle,
\label{eq:inv1_decom} \\
\ab{2} &= \langle0,1\rangle^2 - 2\langle1,1\rangle
  + \langle2,0\rangle, \\
\ab{3} &= -\langle0,1\rangle^3 + 3\langle0,1\rangle \langle1,1\rangle
  -3\langle2,1\rangle + \langle3,0\rangle, \\
\ab{4} &= \langle 0,1\rangle^4 - 4\langle0,1\rangle^2\langle1,1\rangle
  + 4\langle0,1\rangle \langle2,1\rangle + 2\langle1,1\rangle^2
  - 4\langle3,1\rangle + \langle4,0\rangle, 
\label{eq:inv4_decom}\\
\ab{5} &= -\langle0,1\rangle^5
  + 5\langle0,1\rangle^3 \langle1,1\rangle
  - 5\langle0,1\rangle^2\langle2,1\rangle
  - 5\langle0,1\rangle \langle1,1\rangle^2
  + 5\langle0,1\rangle \langle3,1\rangle
  + 5\langle1,1\rangle \langle2,1\rangle \non
&\phantom{=\ }
  - 5\langle4,1\rangle
  + \langle5,0\rangle.
\label{eq:inv5_decom}\\
\ab{6} &= \langle0,1\rangle^6
  - 6\langle0,1\rangle^4\langle1,1\rangle
  + 6\langle0,1\rangle^3\langle2,1\rangle
  + 9\langle0,1\rangle^2\langle1,1\rangle^2
  - 6\langle0,1\rangle^2\langle3,1\rangle \non
&\phantom{=\ }
  - 12\langle0,1\rangle\langle2,1\rangle\langle1,1\rangle
  - 2\langle1,1\rangle^3
  + 6\langle0,1\rangle\langle4,1\rangle
  + 6\langle3,1\rangle\langle1,1\rangle
  + 3\langle2,1\rangle^2
  - 6\langle5,1\rangle \non
&\phantom{=\ }
  + \langle6,0\rangle.
\label{eq:inv6_decom}
\end{align}

\subsection{2, 4 and 6 derivatives}

As a warm-up, let us first consider the Hamiltonians for the
generalized Skyrme model, i.e.~for the Lagrangian with the kinetic
term, the Skyrme term and the BPS-Skyrme
term \cite{Adam:2010fg,Adam:2010ds,Gudnason:2013qba,Gudnason:2014jga,Gudnason:2014hsa,Gudnason:2015nxa,Gudnason:2016yix}. 
Writing the Lagrangians in terms of the time-dependent brackets, we
get 
\begin{align}
\mathcal{L}_2 &= c_{2|2} \left(
  \langle 0,1\rangle - \langle 1,0\rangle
  \right),\\
\mathcal{L}_4 &= \frac{c_{4|2,2}}{2} \left(
  2\langle 0,1\rangle\langle 1,0\rangle
  - 2\langle 1,1\rangle
  + \langle 2,0\rangle
  - \langle 1,0\rangle^2 
  \right),\\
\mathcal{L}_6 &= \frac{c_{6|2,2,2}}{3} \bigg(
  -\frac{3}{2}\langle 0,1\rangle\langle 2,0\rangle
  +\frac{3}{2}\langle 0,1\rangle\langle 1,0\rangle^2
  +3\langle 2,1\rangle
  -3\langle 1,1\rangle\langle 1,0\rangle\non
&\phantom{=c_{6|2,2,2} \bigg(\ }
  -\langle 3,0\rangle
  +\frac{3}{2}\langle 1,0\rangle\langle 2,0\rangle 
  -\frac{1}{2}\langle 1,0\rangle^3
  \bigg),
\end{align}
from which we can calculate the conjugate momenta
\begin{align}
\boldsymbol{\pi}^{(2)}\cdot\mathbf{n}_0 &= 2c_{2|2} \langle 0,1\rangle,\\
\boldsymbol{\pi}^{(4)}\cdot\mathbf{n}_0 &= 2c_{4|2,2} \left(
  \langle 0,1\rangle\langle 1,0\rangle - \langle 1,1\rangle
  \right),\\
\boldsymbol{\pi}^{(6)}\cdot\mathbf{n}_0 &= c_{6|2,2,2}\left(
  -\langle 0,1\rangle\langle 2,0\rangle
  +\langle 0,1\rangle\langle 1,0\rangle^2
  +2\langle 2,1\rangle
  -2\langle 1,1\rangle\langle 1,0\rangle\right).
\end{align}
We can now write down the Hamiltonians in terms of the invariants with
the time-dependent brackets
\begin{align}
\mathcal{H}_2 &= c_{2|2}\left(
  \langle 0,1\rangle + \langle 1,0\rangle\right),\\
\mathcal{H}_4 &= \frac{c_{4|2,2}}{2}\left(
  2\langle 0,1\rangle\langle 1,0\rangle
  -2\langle 1,1\rangle
  -\langle 2,0\rangle
  +\langle 1,0\rangle^2
  \right),\\
\mathcal{H}_6 &= \frac{c_{6|2,2,2}}{3}\bigg(
  -\frac{3}{2}\langle 0,1\rangle\langle 2,0\rangle
  +\frac{3}{2}\langle 0,1\rangle\langle 1,0\rangle^2
  +3\langle 2,1\rangle
  -3\langle 1,1\rangle\langle 1,0\rangle \non
&\phantom{= c_{6|2,2,2}\bigg(\ }
  +\langle 3,0\rangle 
  -\frac{3}{2}\langle 1,0\rangle\langle 2,0\rangle
  +\frac{1}{2}\langle 1,0\rangle^3
  \bigg).
\end{align}
From the invariants, it is not clear whether the Hamiltonians are
bounded from below or not.
Therefore, it is convenient to rewrite them in terms of the
eigenvalues $\lambda_\mu$,
\begin{align}
\mathcal{H}_2 &= c_{2|2}\left(
  \lambda_0^2 + \lambda_1^2 + \lambda_2^2 + \lambda_3^2
  \right),
\label{eq:H2}\\
\mathcal{H}_4 &= c_{4|2,2}\left(
  \lambda_0^2\left(\lambda_1^2 + \lambda_2^2 + \lambda_3^2\right)
  +\lambda_1^2\lambda_2^2
  +\lambda_1^2\lambda_3^2
  +\lambda_2^2\lambda_3^2
  \right),\\
\mathcal{H}_6 &= c_{6|2,2,2}\left(
  \lambda_0^2\left(\lambda_1^2\lambda_2^2 + \lambda_1^2\lambda_3^2
    +\lambda_2^2\lambda_3^2\right)
  +\lambda_1^2\lambda_2^2\lambda_3^2
  \right),
\label{eq:H6}
\end{align}
where we have used the eigenvalues, $\lambda_\mu$, defined as
\begin{equation}
\widetilde{D}_{\mu\nu} \equiv (\mathbf{n}_\mu\cdot\mathbf{n}_\nu) = 
\left[
\widetilde{V}
\begin{pmatrix}
\lambda_0^2\\
&\lambda_1^2\\
&&\lambda_2^2\\
&&&\lambda_3^2
\end{pmatrix}
\widetilde{V}^{\rm T}
\right]_{\mu\nu}, \qquad
\widetilde{V} \equiv
\begin{pmatrix}
\sigma & w^{\rm T}\\
u & V
\end{pmatrix},
\label{eq:eigenvalues_mu}
\end{equation}
where $\sigma$ is a real scalar, $v,w$ are real row-vectors of length
3 and $V$ is a 3-by-3 real matrix.
$\widetilde{V}\widetilde{V}^{\rm T}=\widetilde{V}^{\rm T}\widetilde{V}=\mathbf{1}_4$, which gives rise to the relations\footnote{
Actually the decomposition into temporal and spatial parts of
$\widetilde{D}_{\mu\nu}$ is not necessary when the Hamiltonian only
contains terms with 2 time derivatives, because in that case, one can
form SO(4) singlets, e.g.~
\begin{align}
2\langle 0,1\rangle\langle 1,0\rangle
  -2\langle 1,1\rangle
  -\langle 2,0\rangle
  +\langle 1,0\rangle^2
&= 2\widetilde{D}_{00} \widetilde{D}_{ii}
  - 2\widetilde{D}_{0i}\widetilde{D}_{i0}
  - \widetilde{D}_{ij}\widetilde{D}_{ji}
  + \widetilde{D}_{ii}^2 
= \widetilde{D}_{\mu\mu}^2
- \widetilde{D}_{\mu\nu}\widetilde{D}_{\nu\mu}.
\end{align}
This will not be the case for more than two time derivatives, as we
will see in the next subsection. }
\begin{align}
\sigma^2 + w^{\rm T} w
= \sigma^2 + u^{\rm T} u = 1, \qquad
w w^{\rm T} + V V^{\rm T} = \mathbf{1}_3, \qquad
u^{\rm T} V = -\sigma w^{\rm T}.
\end{align}

We can clearly see that all the Hamiltonians
(Eqs.~(\ref{eq:H2}-\ref{eq:H6})) are positive definite even when
including time dependence. 

It is easy to show that the determinant of the matrix
$\widetilde{D}_{\mu\nu}$ vanishes. This can be checked explicitly by
using a parametrization of $n$ with manifest unit length,
e.g.~$\mathbf{n}=(\sin f\sin g\sin h, \sin f\sin g\cos h, \sin f\cos
g, \cos f)$.
Alternatively this can be understood by noting that the target space
is three dimensional and there are no four independent tangent vectors
$\mathbf{n}_\mu$ which in turn implies that the determinant of
$\widetilde{D}$ vanishes (because one of the vectors must be linearly
dependent on the others)\footnote{We thank Martin Speight for pointing
this out. }.
This has the following implication: one can always choose
$\lambda_0=0$.
This simplifies the Hamiltonians to
\begin{align}
\mathcal{H}_2 &= c_{2|2}\left(
  \lambda_1^2 + \lambda_2^2 + \lambda_3^2
  \right),
\label{eq:H2simpl}\\
\mathcal{H}_4 &= c_{4|2,2}\left(
  \lambda_1^2\lambda_2^2
  +\lambda_1^2\lambda_3^2
  +\lambda_2^2\lambda_3^2
  \right),\\
\mathcal{H}_6 &= c_{6|2,2,2} \lambda_1^2\lambda_2^2\lambda_3^2.
\label{eq:H6simpl}
\end{align}
Obviously, all three Hamiltonians are positive semi-definite.

In the following subsections, we will check explicitly whether
it also possible to establish positivity also the higher-order
Lagrangians constructed in the previous section.

\subsection{8 derivatives}

Let us first rewrite the Lagrangian \eqref{eq:L8rdelta4general} in
terms of the time-dependent brackets defined in
Eq.~\eqref{eq:angular_brackets} using
Eqs.~(\ref{eq:inv1_decom}-\ref{eq:inv4_decom})
\begin{align}
\mathcal{L}_8 &=
\frac{1}{2}\left(4a_4 + 3a_{3,1}\right) \ab{0,1}^2\ab{2,0}
-\frac{1}{2}\left(8a_4 + 3a_{3,1} + 8a_{2,2}\right) \ab{0,1}^2\ab{1,0}^2
+ a_{3,1} \ab{0,1}\ab{3,0}
\non&\phantom{=\ }
-(4a_4 + 3a_{3,1} + 4a_{2,2}) \ab{0,1}\ab{2,0}\ab{1,0}
+(4a_4 + 2a_{3,1} + 4a_{2,2}) \ab{0,1}\ab{1,0}^3
\non&\phantom{=\ }
-(4a_4 + 3a_{3,1}) \ab{0,1}\ab{2,1}
+(8a_4 + 3a_{3,1} + 8a_{2,2}) \ab{0,1}\ab{1,1}\ab{1,0}
+ 4a_4 \ab{3,1}
\non&\phantom{=\ }
+ 3a_{3,1} \ab{2,1}\ab{1,0}
- 2(a_4 + 2a_{2,2}) \ab{1,1}^2
+ 4a_{2,2} \ab{1,1}\ab{2,0}
\non&\phantom{=\ }
- (4a_4 + 3a_{3,1} + 4a_{2,2}) \ab{1,1}\ab{1,0}^2
-a_4 \ab{4,0}
-a_{3,1} \ab{3,0}\ab{1,0}
-a_{2,2} \ab{2,0}^2
\non&\phantom{=\ }
+\frac{1}{2}(4a_4 + 3a_{3,1} + 4a_{2,2}) \ab{2,0}\ab{1,0}^2
-\frac{1}{2}(2a_4 + a_{3,1} + 2a_{2,2}) \ab{1,0}^4.
\end{align}
The conjugate momentum can thus readily be obtained as
\begin{align}
\frac{1}{2}\boldsymbol{\pi}^{(8)}\cdot\mathbf{n}_0 &=
\left(4a_4 + 3a_{3,1}\right) \ab{0,1}^2\ab{2,0}
-\left(8a_4 + 3a_{3,1} + 8a_{2,2}\right) \ab{0,1}^2\ab{1,0}^2
+ a_{3,1} \ab{0,1}\ab{3,0}
\non&\phantom{=\ }
-(4a_4 + 3a_{3,1} + 4a_{2,2}) \ab{0,1}\ab{2,0}\ab{1,0}
+(4a_4 + 2a_{3,1} + 4a_{2,2}) \ab{0,1}\ab{1,0}^3
\non&\phantom{=\ }
-2(4a_4 + 3a_{3,1}) \ab{0,1}\ab{2,1}
+2(8a_4 + 3a_{3,1} + 8a_{2,2}) \ab{0,1}\ab{1,1}\ab{1,0}
+ 4a_4 \ab{3,1}
\non&\phantom{=\ }
+ 3a_{3,1} \ab{2,1}\ab{1,0}
- 4(a_4 + 2a_{2,2}) \ab{1,1}^2
+ 4a_{2,2} \ab{1,1}\ab{2,0}
\non&\phantom{=\ }
- (4a_4 + 3a_{3,1} + 4a_{2,2}) \ab{1,1}\ab{1,0}^2.
\end{align}
It is now straightforward to get the Hamiltonian
\begin{align}
\mathcal{H}_8 &=
\frac{3}{2}\left(4a_4 + 3a_{3,1}\right) \ab{0,1}^2\ab{2,0}
-\frac{3}{2}\left(8a_4 + 3a_{3,1} + 8a_{2,2}\right) \ab{0,1}^2\ab{1,0}^2
+ a_{3,1} \ab{0,1}\ab{3,0}
\non&\phantom{=\ }
-(4a_4 + 3a_{3,1} + 4a_{2,2}) \ab{0,1}\ab{2,0}\ab{1,0}
+(4a_4 + 2a_{3,1} + 4a_{2,2}) \ab{0,1}\ab{1,0}^3
\non&\phantom{=\ }
-3(4a_4 + 3a_{3,1}) \ab{0,1}\ab{2,1}
+3(8a_4 + 3a_{3,1} + 8a_{2,2}) \ab{0,1}\ab{1,1}\ab{1,0}
+ 4a_4 \ab{3,1}
\non&\phantom{=\ }
+ 3a_{3,1} \ab{2,1}\ab{1,0}
- 6(a_4 + 2a_{2,2}) \ab{1,1}^2
+ 4a_{2,2} \ab{1,1}\ab{2,0}
\non&\phantom{=\ }
- (4a_4 + 3a_{3,1} + 4a_{2,2}) \ab{1,1}\ab{1,0}^2
+a_4 \ab{4,0}
+a_{3,1} \ab{3,0}\ab{1,0}
+a_{2,2} \ab{2,0}^2
\non&\phantom{=\ }
-\frac{1}{2}(4a_4 + 3a_{3,1} + 4a_{2,2}) \ab{2,0}\ab{1,0}^2
+\frac{1}{2}(2a_4 + a_{3,1} + 2a_{2,2}) \ab{1,0}^4.
\end{align}
The final step is thus to rewrite the invariants in terms of the
eigenvalues $\lambda_\mu$ using Eq.~\eqref{eq:eigenvalues_mu},
\begin{align}
\mathcal{H}_8 &=
c_{8|4,4}\left(\lambda_1^4\lambda_2^4 + \lambda_1^4\lambda_3^4
  + \lambda_2^4\lambda_3^4\right)
+c_{8|4,2,2}\left(
  \lambda_1^4\lambda_2^2\lambda_3^2
  +\lambda_1^2\lambda_2^4\lambda_3^2
  +\lambda_1^2\lambda_2^2\lambda_3^4\right)
\non&\phantom{=\ }
-4c_{8|4,4}\left(\lambda_1^4 + \lambda_2^4 + \lambda_3^4\right)
  \left[w^{\rm T}\lambda_i^2w\right]^2
-4c_{8|4,2,2}\left(\lambda_1^2\lambda_2^2
  +\lambda_1^2\lambda_3^2 +\lambda_2^2\lambda_3^2\right)
  \left[w^{\rm T}\lambda_i^2w\right]^2
\non&\phantom{=\ }
+4c_{8|4,2,2}\left(\lambda_1^2 + \lambda_2^2 + \lambda_3^2\right)
  \left[w^{\rm T}(\lambda_i^2)^2w\right]
  \left[w^{\rm T}\lambda_i^2w\right]
\non&\phantom{=\ }
-4c_{8|4,4}\left[w^{\rm T}(\lambda_i^2)^2w\right]^2
+4(2c_{8|4,4}-c_{8|4,2,2})\left[w^{\rm T}(\lambda_i^2)^3w\right]
  \left[w^{\rm T}\lambda_i^2w\right],
\end{align}
where the coefficients $c_{8|4,4}$ and $c_{8|4,2,2}$ are defined in
Eq.~\eqref{eq:c8gen_coeff} and
$\lambda_i^2=\diag(\lambda_1^2,\lambda_2^2,\lambda_3^2)$ is the 3-by-3
diagonal matrix of eigenvalues.

Note that the following inner products are positive semi-definite
\beq
w^{\rm T}(\lambda_i^2)^p w \geq 0, \qquad p\in\mathbb{Z}_{>0},
\eeq
as are the eigenvalues themselves, $\lambda_\mu^2\geq 0$, with $\mu$
not summed over.
Writing out explicitly the above inner product, we get 
\beq
w^{\rm T} (\lambda_i^2)^p w
= w_1^2\lambda_1^{2p} + w_2^2\lambda_2^{2p} + w_3^2\lambda_3^{2p}.
\label{eq:wlambdaw}
\eeq
Plugging this into the Hamiltonian, we can write
\begin{align}
\mathcal{H}_8 &=
c_{8|4,4}\left(1 - 4(w_1^2+w_2^2)\right) \llij1244
+c_{8|4,4}\left(1 - 4(w_1^2+w_3^2)\right) \llij1344
\non&\phantom{=\ }
+c_{8|4,4}\left(1 - 4(w_2^2+w_3^2)\right) \llij2344
\non&\phantom{=\ }
+\left[c_{8|4,2,2}\left(1 - 4w_1^2(w^{\rm T} w)\right)
  - 8c_{8|4,4} w_2^2w_3^2\right] \llls422
\non&\phantom{=\ }
+\left[c_{8|4,2,2}\left(1 - 4w_2^2(w^{\rm T} w)\right)
  - 8c_{8|4,4} w_1^2w_3^2\right] \llls242
\non&\phantom{=\ }
+\left[c_{8|4,2,2}\left(1 - 4w_3^2(w^{\rm T} w)\right)
  - 8c_{8|4,4} w_1^2w_2^2\right] \llls224,
\label{eq:H8lambda}
\end{align}
from which it is easy to read off when the instability kicks in.
Since $\sigma^2+w^{\rm T}w=1$, the length of $w$ cannot exceed 1, but
that is not sufficient to establish stability of the Hamiltonian.

It is also clear from the above expression that as long as $w$ is
small enough, the Hamiltonian is positive definite (for any values of
$\lambda_i^2$).

There are two sources of minus signs in the calculation of the
Hamiltonian; one comes from the fact that the square of the time-time
component of the inverse metric is not negative.
The second-order time derivatives in the Lagrangian density are
accompanied by 1 factor of the inverse metric giving exactly 1 minus
sign and hence that term is positive in the Lagrangian and also in the
Hamiltonian.
The fourth-order time derivatives, however, are accompanied by two
factors of the inverse metric giving a plus and hence the term becomes
negative both in the Lagrangian and Hamiltonian.\footnote{Recall that
we use the mostly-positive metric signature. The conclusion remains
the same by using the mostly-negative metric signature, although the
details change. }
The other source of minus signs comes from our desire to eliminate
higher powers of derivatives in the $i$-th direction.

Throughout the paper, we have only used the
invariants \eqref{eq:nninv}.
However, we mentioned another time-dependent
invariant \eqref{eq:epsilon_inv}, which we neglected so far because it
vanishes for static configurations.
Since we are considering time-dependent perturbations in this section,
we should consider including it.
By construction it has 4 derivatives, but each derivative appears only
once in each direction.
We choose to impose parity and time-reversion symmetry on the
Lagrangian, which implies that the invariant \eqref{eq:epsilon_inv}
can only appear with even powers.
Hence, the first Lagrangian where it can appear (squared) is the
eighth-order Lagrangian discussed in this section.
Let us calculate its contribution explicitly
\begin{align}
\mathcal{L}_8' &=
\frac{a_\epsilon}{144}
\left(\epsilon_{abcd}\epsilon^{\mu\nu\rho\sigma}
n_\mu^a n_\nu^b n_\rho^c n_\sigma^d
\right)^2 \non
&= a_{\epsilon} \left(
-\ab{4} + \frac{4}{3}\ab{3}\ab{1} + \frac{1}{2}\ab{2}^2
- \ab{2}\ab{1}^2 + \frac{1}{6}\ab{1}^4
\right).
\label{eq:Linstanton_invariant}
\end{align}
We claimed that the invariant vanishes for static contributions and so
should its square; we can confirm this statement explicitly by
observing that static part of the above Lagrangian is exactly
Eq.~\eqref{eq:nontrivial4rel} and the claim follows.
Turning on time-dependence, the above Lagrangian can be written in
terms of time-dependent brackets in Eq.~\eqref{eq:angular_brackets}
using  Eqs.~(\ref{eq:inv1_decom}-\ref{eq:inv4_decom}) as
\begin{align}
\mathcal{L}_8' &= a_\epsilon \bigg(
  -\frac{4}{3} \ab{0,1}\ab{3,0}
  +2 \ab{0,1}\ab{2,0}\ab{1,0}
  -\frac{2}{3} \ab{0,1}\ab{1,0}^3
  +4 \ab{3,1}
  -4 \ab{2,1}\ab{1,0}
\non&\phantom{=a_\epsilon(\ \ }
  -2 \ab{1,1}\ab{2,0}
  +2 \ab{1,1}\ab{1,0}^2
 -\ab{4,0}
 +\frac{4}{3} \ab{3,0}\ab{1,0}
 +\frac{1}{2} \ab{2,0}^2
 - \ab{2,0}\ab{1,0}^2
\non&\phantom{=a_\epsilon(\ \ }
 +\frac{1}{6} \ab{1,0}^4
\bigg).
\end{align}
We now want to perform a Legendre transformation to get the
corresponding Hamiltonian, starting with writing down the conjugate
momenta
\begin{align}
\frac{1}{2}\boldsymbol{\pi}^{(8)}{}'\cdot\mathbf{n}_0 &=
a_\epsilon \bigg(
  -\frac{4}{3} \ab{0,1}\ab{3,0}
  +2 \ab{0,1}\ab{2,0}\ab{1,0}
  -\frac{2}{3} \ab{0,1}\ab{1,0}^3
  +4 \ab{3,1}
  -4 \ab{2,1}\ab{1,0}
\non&\phantom{=a_\epsilon(\ \ }
  -2 \ab{1,1}\ab{2,0}
  +2 \ab{1,1}\ab{1,0}^2
\bigg),
\end{align}
and hence the Hamiltonian is simply
\begin{align}
\mathcal{H}_8' &=
a_\epsilon \bigg(
  -\frac{4}{3} \ab{0,1}\ab{3,0}
  +2 \ab{0,1}\ab{2,0}\ab{1,0}
  -\frac{2}{3} \ab{0,1}\ab{1,0}^3
  +4 \ab{3,1}
  -4 \ab{2,1}\ab{1,0}
\non&\phantom{=a_\epsilon(\ \ }
  -2 \ab{1,1}\ab{2,0}
  +2 \ab{1,1}\ab{1,0}^2
 +\ab{4,0}
 -\frac{4}{3} \ab{3,0}\ab{1,0}
 -\frac{1}{2} \ab{2,0}^2
 + \ab{2,0}\ab{1,0}^2
\non&\phantom{=a_\epsilon(\ \ }
 -\frac{1}{6} \ab{1,0}^4
\bigg).
\end{align}
Rewriting it in terms of the eigenvalues $\lambda_\mu$ using
Eq.~\eqref{eq:eigenvalues_mu}, we get
\begin{align}
\mathcal{H}_8' &= -4a_\epsilon
  \lambda_0^2\lambda_1^2\lambda_2^2\lambda_3^2 = 0.
\end{align}
As discussed in the previous subsection, one of the eigenvalues
$\lambda_\mu$ must vanish and we can always choose it to be
$\lambda_0$. 
In any case, the above contribution vanishes identically.

We have seen in this subsection that although the static energy of the
Lagrangian \eqref{eq:L8rdelta4general} is positive definite, the
total energy obtained from the corresponding Hamiltonian is not.
Thus the energy is not bounded from below and in principle the theory
is unstable. 

Two comments are in store on this account.
The dynamical instability encountered here is not exactly
due to Ostrogradsky's theorem \cite{Woodard:2015zca}, because our
Lagrangian by construction (by choice) does not contain $\square n^a$,
which requires a second conjugate momentum for the field $n^a$, see
also App.~\ref{app:diffOstrogradsky}.
To flesh this point out in more details, let us write the conjugate
momenta $\boldsymbol{\pi}^{(8)}$ in details before dotting them onto
$\mathbf{n}_0$ as
\begin{align}
\pi^{(8)a} &=
2\left(4a_4 + 3a_{3,1}\right) \ab{0,1}\ab{2,0} n_0^a
-2\left(8a_4 + 3a_{3,1} + 8a_{2,2}\right) \ab{0,1}\ab{1,0}^2 n_0^a
+ 2a_{3,1} \ab{3,0} n_0^a
\non&\phantom{=\ }
-2(4a_4 + 3a_{3,1} + 4a_{2,2}) \ab{2,0}\ab{1,0} n_0^a
+2(4a_4 + 2a_{3,1} + 4a_{2,2}) \ab{1,0}^3 n_0^a
\non&\phantom{=\ }
-2(4a_4 + 3a_{3,1})
  \left(\ab{2,1} n_0^a
    + \ab{0,1} n_i^a (\mathbf{n}_i\cdot\mathbf{n}_j)
    (\mathbf{n}_j\cdot\mathbf{n}_0)\right) 
\non&\phantom{=\ }
+2(8a_4 + 3a_{3,1} + 8a_{2,2})
  \left(
  \ab{1,1}\ab{1,0} n_0^a
  +\ab{0,1}\ab{1,0} n_i^a (\mathbf{n}_i\cdot\mathbf{n}_0) 
  \right)
\non&\phantom{=\ }
+ 8a_4 n_i^a (\mathbf{n}_i\cdot\mathbf{n}_j) 
  (\mathbf{n}_j\cdot\mathbf{n}_k) (\mathbf{n}_k\cdot\mathbf{n}_0) 
+ 6a_{3,1} \ab{1,0} n_i^a (\mathbf{n}_i\cdot\mathbf{n}_j)
    (\mathbf{n}_j\cdot\mathbf{n}_0)
\non&\phantom{=\ }
- 8(a_4 + 2a_{2,2}) \ab{1,1} n_i^a (\mathbf{n}_i\cdot\mathbf{n}_0)
+ 8a_{2,2} \ab{2,0} n_i^a (\mathbf{n}_i\cdot\mathbf{n}_0)
\non&\phantom{=\ }
- 2(4a_4 + 3a_{3,1} + 4a_{2,2}) \ab{1,0}^2
  n_i^a (\mathbf{n}_i\cdot\mathbf{n}_0).
\end{align}
Notice that we can write the conjugate momenta as
\begin{align}
\pi^{(8)a} &=
2\left(K_0^{ab} + \ab{0,1} K_1^{ab} + \ab{1,1} K_2^{ab}
  + \ab{2,1} K_3^{ab} \right)n_0^b \non
&\phantom{=\ }
+ 2\left(K_0 + \ab{0,1} K_1 + K_2 \ab{1,1}\right)\delta^{ab} n_0^b.
\end{align}
In principle, now we would like to invert the equation to get an
expression for $n_0^a$ in terms of $\pi^{(8)a}$.
The equation, however, is a cubic matrix equation; we will not attempt
at finding the explicit solution here.
It is merely enough to notice that the inverse, which we assume to
exist, is proportional to a cubic root involving $\pi^{(8)a}$ itself. 
Therefore, the Hamiltonian does not contain a term linear in
$\boldsymbol{\pi}$ (which does not appear anywhere else in the
Hamiltonian) and the Ostrogradsky theorem hence does not apply. 
The instability is thus much more intricate and of nonlinear nature
than the Ostrogradsky one. 

Our theory is a highly nonlinear field theory and the dynamical
instability is rooted in this nonlinearity. 
In fact there are two different effects destabilizing the Hamiltonian
at hand.
The first is due to the Lagrangian being composed of products of
Lorentz invariants. When a term contains four time derivatives it is
necessarily accompanied by two inverse metric factors, thus giving the
same sign as for the potential part of the Lagrangian.
This induces a ghost-like kinetic (squared) term in the Hamiltonian,
which thus is not bounded from below.
Clearly this effect occurs for all even powers of squared time
derivatives, but not for odd powers (like 2,6,10 and so on).
A different effect destabilizing the system is due to higher powers
(than two) of time derivatives giving larger factors in the conjugate
momentum (and also in the Euler-Lagrange equations of motion of
course) and this in turn implies that the Hamiltonian does not
recombine Lorentz SO(3,1) invariants as SO(4) invariants; this SO(4)
invariance is broken and that is why $w$ appears in the
result \eqref{eq:H8lambda}. 
This yields mixed terms of both signs; of course the reason for the
mixed terms of both sign is that we used constraints to obtain a
minimal $\delta=4$ Lagrangian. After breaking the would-be SO(4)
symmetry of the terms in the Hamiltonian, these constraints induce
terms of both signs.

Even though the Hamiltonian \eqref{eq:H8lambda} is not positive
definite, it clearly provides conditions for stability.
If all factors in front of the $\lambda$s are positive, then the
system is stable at the time-dependent level.
This can be achieved in different ways; for instance, we could choose
$c_{8|4,4}=0$, $c_{8|4,2,2}>0$ and require the following condition 
\beq
w_i^2(w^{\rm T}w) < \frac{1}{4}, \quad \forall i\in(1,2,3).
\eeq
For $c_{8|4,4}>0$ additional constraints are required to retain a
positive definite Hamiltonian. 
It is also clear what the physical meaning of the above constraint is;
in the static limit $w=0$ and so $w$ is a vector that rotates the
time-dependence of the strain tensor $\widetilde{D}_{\mu\nu}$ into the
nonvanishing eigenvalues $\lambda_i^2$.

We will show this more explicitly with an example in the next section.
In the following subsections, however, we will continue with the
minimal $\delta=4$ Lagrangians and check that what we observed for the
eighth-order Lagrangian is general and thus persists for the
tenth-order and twelfth-order Lagrangians.

\subsection{10 derivatives}

We will now calculate the Hamiltonian corresponding to the
Lagrangian \eqref{eq:L10geninv_nonsimpl} along the lines of the last
subsection.
Since the calculation is mostly mechanical and we showed the explicit
calculations for the eighth-order Lagrangian in the last subsection,
we will not flesh out the steps here, but simply state the result
\begin{align}
\mathcal{H}_{10} &=
 (5a_5 + 4a_{4,1}) \ab{0,1}^2\ab{3,0}
-3(5a_5 + 2a_{4,1} + 3a_{3,2}) \ab{0,1}^2\ab{2,0}\ab{1,0}
\non&\phantom{=\ }
+(10a_5 + 2a_{4,1} + 9a_{3,2}) \ab{0,1}^2\ab{1,0}^3 
+a_{4,1} \ab{0,1}\ab{4,0}
\non&\phantom{=\ }
-\frac{2}{3} (5a_5 + 4a_{4,1} + 3a_{3,2}) \ab{0,1}\ab{3,0}\ab{1,0}
-\frac{1}{4} (5a_5 + 2a_{4,1} + 6a_{3,2}) \ab{0,1}\ab{2,0}^2
\non&\phantom{=\ }
+\frac{1}{2}\left(15a_5 + 6a_{4,1} + 12a_{3,2}\right) \ab{0,1}\ab{2,0}\ab{1,0}^2
-\frac{5}{12} (7a_5 + 2a_{4,1} + 6a_{3,2}) \ab{0,1}\ab{1,0}^4
\non&\phantom{=\ }
-3(5a_5 + 4a_{4,1}) \ab{0,1}\ab{3,1}
+3(10a_5 + 4a_{4,1} + 6a_{3,2}) \ab{0,1}\ab{2,1}\ab{1,0}
\non&\phantom{=\ }
+3(5a_5 + 2a_{4,1} + 3a_{3,2}) \ab{0,1}\ab{1,1}\ab{2,0}
-3(10a_5 + 2a_{4,1} + 9a_{3,2}) \ab{0,1}\ab{1,1}\ab{1,0}^2
\non&\phantom{=\ }
+5 a_5 \ab{4,1}
+4 a_{4,1} \ab{3,1}\ab{1,0}
-3(5a_5 + 6a_{3,2}) \ab{2,1}\ab{1,1}
+ 3 a_{3,2} \ab{2,1}\ab{2,0}
\non&\phantom{=\ }
-(5a_5 + 4a_{4,1} + 3a_{3,2}) \ab{2,1}\ab{1,0}^2
+3(5a_5 + 6a_{3,2}) \ab{1,1}^2\ab{1,0}
+2 a_{3,2} \ab{1,1}\ab{3,0}
\non&\phantom{=\ }
-(5a_5 + 2a_{4,1} + 6a_{3,2}) \ab{1,1}\ab{2,0}\ab{1,0}
+(5a_5 + 2a_{4,1} + 4a_{3,2}) \ab{1,1}\ab{1,0}^3
-a_5 \ab{5,0}
\non&\phantom{=\ }
-a_{4,1} \ab{4,0}\ab{1,0}
-a_{3,2} \ab{3,0}\ab{2,0}
+\frac{1}{3}(a_{3,2} + 4 a_{4,1} + 5 a_5) \ab{3,0}\ab{1,0}^2
\non&\phantom{=\ }
+\frac{1}{4} (6 a_{3,2} + 2 a_{4,1} + 5 a_5) \ab{2,0}^2\ab{1,0}
-\frac{1}{2} (4 a_{3,2} - 2 a_{4,1} - 5 a_5) \ab{2,0}\ab{1,0}^3
\non&\phantom{=\ }
+\frac{1}{12} (6 a_{3,2} + 2 a_{4,1} + 7 a_5) \ab{1,0}^5.
\end{align}
Rewriting the Hamiltonian in terms of the eigenvalues, $\lambda_\mu$,
using Eqs.~\eqref{eq:eigenvalues_mu} and \eqref{eq:wlambdaw}, we get 
\begin{align}
\mathcal{H}_{10} &=
c_{10|4,4,2}\left(1 - 4w_1^2 - 4w_2^2 - 8w_1^2w_2^2 - 4w_1^2w_3^2 -
4w_2^2w_3^2\right) \llls442
\non&\phantom{=\ }
+c_{10|4,4,2}\left(1 - 4w_1^2 - 4w_3^2 - 4w_1^2w_2^2 - 8w_1^2w_3^2 -
4w_2^2w_3^2\right) \llls424
\non&\phantom{=\ }
+c_{10|4,4,2}\left(1 - 4w_2^2 - 4w_3^2 - 4w_1^2w_2^2 - 4w_1^2w_3^2 -
8w_2^2w_3^2\right) \llls244.
\label{eq:H10lambda}
\end{align}
Unfortunately, the Hamiltonian is not positive definite for arbitrary
vectors $w$.
The conditions for stability are clear however, viz.~as long as $w$
is small enough the Hamiltonian is positive.

\subsection{12 derivatives}

We will now calculate the Hamiltonian corresponding to the
Lagrangian \eqref{eq:L12geninv_nonsimpl} along the lines of the last
subsection.

A difference with respect to the other cases, however, is that some of
the free parameters in the static energy give rise to terms with 6
time derivatives. 
To eliminate these we set
\beq
a_{4,1,1} = - \frac{3}{2}a_6 - \frac{5}{4}a_{5,1} - a_{4,2},
\eeq
which leaves us with four free parameters $a_6$, $a_{5,1}$, $a_{4,2}$
and $a_{3,3}$. 
The Hamiltonian in terms of the time-dependent brackets reads
\beq
\mathcal{H}_{12} =
\mathcal{H}_{12}^{a}
+\mathcal{H}_{12}^{b}
+\mathcal{H}_{12}^{c},
\eeq
where we have defined
\begin{align}
\mathcal{H}_{12}^{a} &\equiv
\frac{3}{4}(6a_6 + 5a_{5,1})
  \ab{0, 1}^2 \ab{4, 0} 
-(12a_6 + 5a_{5,1} + 8a_{4,2})
  \ab{0, 1}^2 \ab{3, 0} \ab{1, 0}
\non&\phantom{=\ }
-\frac{3}{8}(12a_6 + 5a_{5,1} + 18a_{3,3})
  \ab{0, 1}^2 \ab{2, 0}^2
\non&\phantom{=\ }
+\frac{3}{4}(24a_6 + 5a_{5,1} + 16a_{4,2} + 18a_{3,3})
  \ab{0, 1}^2 \ab{2, 0} \ab{1, 0}^2
\non&\phantom{=\ }
-\frac{1}{8}(48a_6 + 5a_{5,1} + 32a_{4,2} + 54a_{3,3})
  \ab{0, 1}^2 \ab{1, 0}^4
+a_{5,1}
  \ab{0, 1} \ab{5, 0}
\non&\phantom{=\ }
-\frac{1}{2}(6a_6 + 5a_{5,1} + 4a_{4,2})
  \ab{0, 1} \ab{4, 0} \ab{1, 0}
\non&\phantom{=\ }
-\frac{1}{6}(12a_6 + 5a_{5,1} + 4a_{4,2} + 3a_{3,3})
  \ab{0, 1} \ab{3, 0} \ab{2, 0}
\non&\phantom{=\ }
+\frac{1}{2}(12a_6 + 5a_{5,1} + 8a_{4,2} + 6a_{3,3})
  \ab{0, 1} \ab{3, 0} \ab{1, 0}^2
\non&\phantom{=\ }
+\frac{1}{4}(18a_6 + 5a_{5,1} + 12a_{4,2} + 18a_{3,3})
  \ab{0, 1} \ab{2, 0}^2 \ab{1, 0}
\non&\phantom{=\ }
-\frac{1}{3}(21a_6 + 5a_{5,1} + 14a_{4,2} + 18a_{3,3})
  \ab{0, 1} \ab{2, 0} \ab{1, 0}^3
\non&\phantom{=\ }
+\frac{1}{4}(6a_6 + a_{5,1} + 4a_{4,2} + 6a_{3,3})
  \ab{0, 1} \ab{1, 0}^5,
\end{align}
\begin{align}
\mathcal{H}_{12}^{b} &\equiv
-3(6a_6 + 5a_{5,1})
  \ab{0, 1} \ab{4, 1} 
+3(12a_6 + 5a_{5,1} + 8a_{4,2})
  \ab{0, 1} \ab{3, 1} \ab{1, 0}
\non&\phantom{=\ }
+\frac{3}{2}(12a_6 + 5a_{5,1} + 18a_{3,3})
  \ab{0, 1} \ab{2, 1} \ab{2, 0}
\non&\phantom{=\ }
-\frac{3}{2}(24a_6 + 5a_{5,1} + 16a_{4,2} + 18a_{3,3})
  \ab{0, 1} \ab{2, 1} \ab{1, 0}^2
\non&\phantom{=\ }
+(12a_6 + 5a_{5,1} + 8a_{4,2})
  \ab{0, 1} \ab{1, 1} \ab{3, 0}
\non&\phantom{=\ }
-\frac{3}{2}(24a_6 + 5a_{5,1} + 16a_{4,2} + 18a_{3,3})
  \ab{0, 1} \ab{1, 1} \ab{2, 0} \ab{1, 0}
\non&\phantom{=\ }
+\frac{1}{2}(48a_6 + 5a_{5,1} + 32a_{4,2} + 54a_{3,3})
  \ab{0, 1} \ab{1, 1} \ab{1, 0}^3 
+6a_6
  \ab{5, 1} 
+5a_{5,1}
  \ab{4, 1} \ab{1, 0}
\non&\phantom{=\ }
+4a_{4,2}
  \ab{3, 1} \ab{2, 0} 
-6(3a_6 + 4a_{4,2})
  \ab{3, 1} \ab{1, 1} 
-(6a_6 + 5a_{5,1} + 4a_{4,2})
  \ab{3, 1} \ab{1, 0}^2
\non&\phantom{=\ }
-9(a_6 + 3a_{3,3})
  \ab{2, 1}^2 
+6(6a_6 + 4a_{4,2} + 9a_{3,3})
  \ab{2, 1} \ab{1, 1} \ab{1, 0} 
+6a_{3,3}
  \ab{2, 1} \ab{3, 0}
\non&\phantom{=\ }
-\frac{1}{2}(12a_6 + 5a_{5,1} + 8a_{4,2} + 18a_{3,3})
  \ab{2, 1} \ab{2, 0} \ab{1, 0}
\non&\phantom{=\ }
+\frac{1}{2}(12a_6 + 5a_{5,1} + 8a_{4,2} + 6a_{3,3})
  \ab{2, 1} \ab{1, 0}^3 
+3(3a_6 + 4a_{4,2})
  \ab{1, 1}^2 \ab{2, 0}
\non&\phantom{=\ }
-3(6a_6 + 4a_{4,2} + 9a_{3,3})
  \ab{1, 1}^2 \ab{1, 0}^2
+2a_{4,2}
  \ab{1, 1} \ab{4, 0}
\non&\phantom{=\ }
-\frac{1}{3}(12a_6 + 5a_{5,1} + 8a_{4,2} + 18a_{3,3})
  \ab{1, 1} \ab{3, 0} \ab{1, 0} 
-\frac{3}{2}(a_6 + a_{4,2})
  \ab{1, 1} \ab{2, 0}^2
\non&\phantom{=\ }
+\frac{1}{2}(18a_6 + 5a_{5,1} + 12a_{4,2} + 18a_{3,3})
  \ab{1, 1} \ab{2, 0} \ab{1, 0}^2
\non&\phantom{=\ }
-\frac{1}{6}(21a_6 + 5a_{5,1} + 14a_{4,2} + 18a_{3,3})
  \ab{1, 1} \ab{1, 0}^4,
\end{align}
\begin{align}
\mathcal{H}_{12}^{c} &\equiv
a_6
  \ab{6, 0}
+a_{5,1}
  \ab{5, 0} \ab{1, 0}
+a_{4,2}
  \ab{4, 0} \ab{2, 0} 
-\frac{1}{4}(6a_6 + 5a_{5,1} + 4a_{4,2})
  \ab{4, 0} \ab{1, 0}^2
\non&\phantom{=\ }
+a_{3,3}
  \ab{3, 0}^2
-\frac{1}{6}(12a_6 + 5a_{5,1} + 8a_{4,2} + 18a_{3,3})
  \ab{3, 0} \ab{2, 0} \ab{1, 0}
\non&\phantom{=\ }
+\frac{1}{6}(12a_6 + 5a_{5,1} + 8a_{4,2} + 6a_{3,3})
  \ab{3, 0} \ab{1, 0}^3
-\frac{1}{4}(a_6 + 2a_{4,2})
  \ab{2, 0}^3
\non&\phantom{=\ }
+\frac{1}{8}(18a_6 + 5a_{5,1} + 12a_{4,2} + 18a_{3,3})
  \ab{2, 0}^2 \ab{1, 0}^2
\non&\phantom{=\ }
-\frac{1}{12}(21a_6 + 5a_{5,1} + 14a_{4,2} + 18a_{3,3})
  \ab{2, 0} \ab{1, 0}^4
\non&\phantom{=\ }
+\frac{1}{24}(6a_6 + a_{5,1} + 4a_{4,2} + 6a_{3,3})
  \ab{1, 0}^6.
\end{align}
Rewriting the Hamiltonian in terms of the eigenvalues, $\lambda_\mu$,
using Eqs.~\eqref{eq:eigenvalues_mu} and \eqref{eq:wlambdaw}, we get 
\begin{align}
\mathcal{H}_{12} &=
c_{12|4,4,4}
\left(1 - 4(w^{\rm T}w) - 8(w_1^2w_2^2 + w_1^2w_3^2 + w_2^2w_3^2)\right)
\llls{4}{4}{4}.
\label{eq:H12lambda}
\end{align}
Unfortunately, the Hamiltonian is not positive definite.
The condition for stability is nevertheless clear; as long as $w$ is
small enough, the Hamiltonian is positive.

\section{Low-energy stability}\label{sec:low_energy_stability}

In this section we argue that if the theory we constructed is regarded
as a low-energy effective theory, then not only the Skyrmions
themselves can only be described at low energies, but perturbations of
them also have to be below the scale of validity of the effective
theory.\footnote{See also e.g.~Ref.~\cite{Donoghue:2017fvm}. }

Let us first note what happens to the 3-vector $w$ in the static
limit. If we pick the time-time component of the strain tensor and set
$\lambda_0=0$, we get
\beq
\mathbf{n}_0\cdot\mathbf{n}_0 = w^{\rm T}(\lambda_i^2)w,
\label{eq:D00=wlambdaw}
\eeq
which vanishes in the static limit and since $\lambda_i^2$ cannot
vanish, then $w=0$ must hold.
When we turn on time dependence, say by a boost, then what happens is
that the 3 eigenvalues $\lambda_i$ receive corrections like
\beq
\lambda_i^2 = \bar{\lambda}_i^2 + v^2 \lambda_i^{'2}
+ \mathcal{O}(v^4), \qquad \forall i,
\eeq
($i$ not summed over) where $\bar{\lambda}_i$ is the static part of
the eigenvalue and in order for the strain tensor to receive a
nonzero time-time component, $w$ must be nonzero.
We also know from the definition of the diagonalization matrices, that
$\sigma^2+w^{\rm T}w=1$ and it follows that the length of $w$ is
smaller than or equal to unity: $w^{\rm T}w\leq 1$.
The same thus holds for each of the components of $w$. 

It should now be clear from Eq.~\eqref{eq:D00=wlambdaw}, that at small
time derivatives corresponding to small velocities or equivalently to
small energy scales of the perturbations, the components of $w\ll 1$. 
This ameliorates the instability and if the perturbations are
sufficiently small, then the instability does not occur.
Nevertheless, the instability can happen at some critical value of the
derivatives, i.e.~in the product of temporal and spatial derivatives. 
Although a theory which is not manifestly stable is not particularly
desirable, this is somewhat expected, because the expansion in
derivatives implicitly corresponds to a low-energy theory where
high-energy states have been integrated out, leaving higher orders
in derivatives as effective operators in the low-energy effective
theory.
In particular, we expect the scale of validity of the low-energy
effective theory to be below the energy scale where the lowest state
has been integrated out. For the Skyrme model with four derivative
terms, this corresponds to the mass of the $\rho$ meson, while for the 
generalized Skyrme model with only the sixth-order derivative term and
the kinetic term, it corresponds instead to the mass of the $\omega$
meson.

The simplest possible perturbations are just excitations of the lowest
lying modes of the spectrum of the Skyrmions. The lowest modes are of
course the zero modes, including the translational moduli (other are
rotational modes etc.).
Other low-lying modes include vibrational
modes, see e.g.~\cite{Halcrow:2015rvz,Adam:2016lir,Halcrow:2016spb}. 

Here we will consider the simplest possible mode to excite, namely the
translational zero mode. As it is a zero mode, the energy of the
perturbation is simply given by the relativistic energy being
$\gamma(v)$ times the rest mass. Therefore the velocity $v$ translates
into an energy scale.
For other types of perturbations, their frequency translates into an
energy scale. 
Let us take the direction of the motion as $x^1$ for which the Lorentz
boost becomes
\beq
x^1 - x_0^1 \to \frac{x^1 - x_0^1 - v t}{\sqrt{1 - v^2}}
\simeq x^1 - x_0^1 - v t,
\eeq
where we have expanded the Lorentz boost in $v$ so it is simply a
Galilean boost. 
We will hence expand the Skyrmion fields in the velocity $v$
as\footnote{If one considers other modes than the translational zero
modes, the expression below would instead become of the form
\beq
n_0^a = i\omega_a n^a,
\eeq
$a$ not summed over; now the energy scale of the perturbation is
directly set by $\omega_a$. 
}
\beq
\mathbf{n}_0 \equiv \p_0\mathbf{n}
= \p_i\mathbf{n}\frac{\p x^i}{\p t}
= \mathbf{n}_i \delta^{i1} v
= \mathbf{n}_1 v.
\eeq
Since $\mathbf{n}_0$ is proportional to $\mathbf{n}_1$ it is clear
that the determinant of the strain tensor $\widetilde{D}$ vanishes and
hence that $\lambda_0$ can be chosen to vanish.
Although we chose the direction of the boost in this case,
it is always possible to write $\mathbf{n}_0$ as a linear combination
of the other three $\mathbf{n}_i$. 

Since we choose $\lambda_0=0$ to be the vanishing eigenvalue,
$(\sigma, u^{\rm T})^{\rm T}$ is the eigenvector corresponding to the
zero eigenvalue.
In this case of the translational zero modes, we know the form of the
strain tensor
\beq
\widetilde{D}_{\mu\nu} =
\begin{pmatrix}
v^2 \mathbf{n}_1\cdot\mathbf{n}_1 & v \mathbf{n}_1\cdot\mathbf{n}_1 &
v \mathbf{n}_1\cdot\mathbf{n}_2 & v \mathbf{n}_1\cdot\mathbf{n}_3 \\
v \mathbf{n}_1\cdot\mathbf{n}_1 & \mathbf{n}_1\cdot\mathbf{n}_1
& \mathbf{n}_1\cdot\mathbf{n}_2 & \mathbf{n}_1\cdot\mathbf{n}_3 \\
v \mathbf{n}_2\cdot\mathbf{n}_1 & \mathbf{n}_2\cdot\mathbf{n}_1
& \mathbf{n}_2\cdot\mathbf{n}_2 & \mathbf{n}_2\cdot\mathbf{n}_3 \\
v \mathbf{n}_3\cdot\mathbf{n}_1 & \mathbf{n}_3\cdot\mathbf{n}_1
& \mathbf{n}_3\cdot\mathbf{n}_2 & \mathbf{n}_3\cdot\mathbf{n}_3
\end{pmatrix},
\eeq
and so the eigenvector corresponding to the vanishing eigenvalue is 
\beq
\sigma = \frac{1}{\sqrt{1+v^2}}, \qquad
u = -\frac{1}{\sqrt{1+v^2}}\begin{pmatrix} v\\ 0\\ 0\end{pmatrix}. 
\eeq
We need to estimate $w$.
Although we cannot determine $w$ exactly, we know that the length of
$w$ equals that of $u$ and that it is related to $\sigma$ and $u$ via
$V$ as $w=-\sigma^{-1}V^{\rm T}u$:
\beq
w^{\rm T}w = \frac{v^2}{1+v^2}, \qquad
w = v
\begin{pmatrix} V_{11} \\ V_{12} \\ V_{13} \end{pmatrix}
= \frac{v}{\sqrt{1+v^2}}
\begin{pmatrix}
\sin\theta\sin\chi\\
\sin\theta\cos\chi\\
\cos\theta
\end{pmatrix},
\label{eq:wparam}
\eeq
where $\theta$ and $\chi$ are functions of spacetime coordinates and
possibly of velocity $v$.

If we try to expand the
Hamiltonians \eqref{eq:H8lambda}, \eqref{eq:H10lambda}
and \eqref{eq:H12lambda} in small velocity $v\ll 1$, we get 
\begin{align}
\mathcal{H}_8 &=
 c_{8|4,4} (\llijbar1244 + \llijbar1344 + \llijbar2344)
 +c_{8|4,2,2} (\lllsbar422 + \lllsbar242 + \lllsbar224)
\non&\phantom{=\ }
-4c_{8|4,4}v^2\left[
  \sin^2(\bar{\theta}) \llijbar1244
  +(\cos^2\bar{\theta} + \sin^2\bar{\chi}\sin^2\bar{\theta}) \llijbar1344
  +(\cos^2\bar{\theta} + \cos^2\bar{\chi}\sin^2\bar{\theta}) \llijbar2344
  \right]
\non&\phantom{=\ }
+c_{8|4,4} v^2 \left[
  \lambda_1^{'2}(\bar{\lambda}_2^2 + \bar{\lambda}_3^2)
  +\lambda_2^{'2}(\bar{\lambda}_1^2 + \bar{\lambda}_3^2)
  +\lambda_3^{'2}(\bar{\lambda}_1^2 + \bar{\lambda}_2^2)
  \right]
\non&\phantom{=\ }
+2c_{8|4,2,2}v^2
  (\lambda_1^{'2} + \lambda_2^{'2} + \lambda_3^{'2})\lllsbar222
\non&\phantom{=\ }
+c_{8|4,2,2}v^2\left[
   \lambda_1^{'2}(\llijbar2342 + \llijbar2324)
  +\lambda_2^{'2}(\llijbar1342 + \llijbar1324)
  +\lambda_3^{'2}(\llijbar1242 + \llijbar1224)
  \right]
\non&\phantom{=\ }
+\mathcal{O}(v^4),
\end{align}
\begin{align}
\mathcal{H}_{10} &=
c_{10|4,4,2}(\lllsbar442 + \lllsbar424 + \lllsbar244)
\non&\phantom{=\ }
-4c_{10|4,4,2} v^2\Big[
  \sin^2(\bar{\theta}) \lllsbar442
  +(\cos^2\bar{\theta} + \sin^2\bar{\chi}\sin^2\bar{\theta}) \lllsbar424
  \non&\qquad\qquad\qquad\quad
  +(\cos^2\bar{\theta} + \cos^2\bar{\chi}\sin^2\bar{\theta}) \lllsbar244
  \Big]
\non&\phantom{=\ }
+2c_{10|4,4,2}v^2 \lllsbar222 \left[
   \lambda_1^{'2}(\bar{\lambda}_2^2 + \bar{\lambda}_3^2)
  +\lambda_2^{'2}(\bar{\lambda}_1^2 + \bar{\lambda}_3^2)
  +\lambda_3^{'2}(\bar{\lambda}_1^2 + \bar{\lambda}_2^2)
  \right]
\non&\phantom{=\ }
+c_{10|4,4,2} v^2 \left[
   \lambda_1^{'2}\llijbar2344
  +\lambda_2^{'2}\llijbar1344
  +\lambda_3^{'2}\llijbar1244
  \right]
\non&\phantom{=\ }
+\mathcal{O}(v^4),\\
\mathcal{H}_{12} &=
c_{12|4,4,4} \lllsbar444
\non&\phantom{=\ }
-4c_{12|4,4,4} v^2 \lllsbar444
+ 2c_{12|4,4,4} v^2 \lllsbar222 \left[
   \lambda_1^{'2}\llijbar2322
  +\lambda_2^{'2}\llijbar1322
  +\lambda_3^{'2}\llijbar1222
  \right]
\non&\phantom{=\ }
+\mathcal{O}(v^4),
\end{align}
where the barred symbols stand for their static value. 
It is very difficult to prove positivity of the leading order terms in
general (for a general perturbation) because they come with both
signs; note that $\lambda_i^{'2}\in\mathbb{R}$ is only real, but not
necessarily positive in general.
However, the full eigenvalues $\lambda_i^2>0$ are always positive
(semi-)definite for each $i$.
On physical grounds we expect the energy to increase by perturbing the
system, so we expect the leading order correction to be positive.

We can do a bit better by focusing on the translational zero mode.
In order to estimate what happens to the eigenvalues for the
translational zero mode, we expand the spacetime strain tensor
$\widetilde{D}_{\mu\nu}$ to second order in $v$ and find the
eigenvalues are modified as
\beq
\lambda_i^2 = V^{\rm T}\left[
\begin{pmatrix}
\mathbf{n}_1\cdot\mathbf{n}_1 & \mathbf{n}_1\cdot\mathbf{n}_2
  & \mathbf{n}_1\cdot\mathbf{n}_3 \\ 
\mathbf{n}_2\cdot\mathbf{n}_1 & \mathbf{n}_2\cdot\mathbf{n}_2
  & \mathbf{n}_2\cdot\mathbf{n}_3 \\ 
\mathbf{n}_3\cdot\mathbf{n}_1 & \mathbf{n}_3\cdot\mathbf{n}_2
  & \mathbf{n}_3\cdot\mathbf{n}_3
\end{pmatrix}
+v^2
\begin{pmatrix}
2\mathbf{n}_1\cdot\mathbf{n}_1 & \mathbf{n}_1\cdot\mathbf{n}_2
  & \mathbf{n}_1\cdot\mathbf{n}_3 \\
\mathbf{n}_2\cdot\mathbf{n}_1 & 0 & 0\\
\mathbf{n}_3\cdot\mathbf{n}_1 & 0 & 0
\end{pmatrix}
\right] V,
\eeq
where
\beq
\bar{\lambda}_i^2 = \lim_{v\to 0} \lambda_i^2, \qquad
v^2 \lambda_i^{'2} = \lambda_i^2 - \bar{\lambda}_i^2.
\eeq
If we now rescale the coordinate $x^1\to x^{'1}=(1+v^2)^{-1}x^1$ and
note that $(1+v^2)^2\simeq 1+2v^2$ to second order in $v$, then we can
write
\beq
\begin{pmatrix}
\lambda_1^2\\
&\lambda_2^2\\
&&\lambda_3^2
\end{pmatrix}
= V^{\rm T}
\begin{pmatrix}
\mathbf{n}_{1'}\cdot\mathbf{n}_{1'}& \mathbf{n}_{1'}\cdot\mathbf{n}_2
  & \mathbf{n}_{1'}\cdot\mathbf{n}_3 \\ 
\mathbf{n}_2\cdot\mathbf{n}_{1'} & \mathbf{n}_2\cdot\mathbf{n}_2
  & \mathbf{n}_2\cdot\mathbf{n}_3 \\ 
\mathbf{n}_3\cdot\mathbf{n}_{1'} & \mathbf{n}_3\cdot\mathbf{n}_2
  & \mathbf{n}_3\cdot\mathbf{n}_3
\end{pmatrix}
V.
\eeq
Although we have written the diagonalization on the same form as for
the static eigenvalues, it is quite nontrivial to estimate the change
in the eigenvalues $\lambda_i^2$; in general the scaling we performed
will affect all eigenvalues and it is hard to even estimate the size
of the change.

In the case of the twelfth order Hamiltonian, $\mathcal{H}_{12}$ of
Eq.~\eqref{eq:H12lambda}, we know that each term has four derivatives
with respect to $x^1$ and hence it is easy to compare the energies as
follows.
The static energy density of the non-boosted system is
\beq
\mathcal{H}_{12}^0 = c_{12|4,4,4}
\bar{\lambda}_1^4\bar{\lambda}_2^4\bar{\lambda}_3^4,
\eeq
while for the Galilean boosted system, we have
\begin{align}
\mathcal{H}_{12}^{\rm boosted} &= c_{12|4,4,4}(1+v^2)^4(1-4v^2)
\bar{\lambda}_1^4\bar{\lambda}_2^4\bar{\lambda}_3^4 \non
&= c_{12|4,4,4}
\bar{\lambda}_1^4\bar{\lambda}_2^4\bar{\lambda}_3^4
+\mathcal{O}(v^4),
\end{align}
that is, to leading order in $v^2$, the is no instability under the
translational zero mode.
In order to determine stability would require a next-to-leading order
calculation which, however, is very difficult.

For the tenth- and eighth-order Hamiltonians, the derivatives in the
spatial directions are not distributed symmetrically for all terms and
therefore it is not possible to compare the scaled system with the
static one, because the scaling we performed breaks isotropy.
The breaking of isotropy can also be seen from the appearance of
$\sin\theta$ and $\sin\chi$ in the above expressions; it corresponds
to part of the diagonalization matrix $V$ that rotates the
3-dimensional strain tensor into a diagonal form. 
We note, however, that the eighth-order Hamiltonian, $\mathcal{H}_8$
of Eq.~\eqref{eq:H8lambda} is positive definite to leading order in
$v^2$ if we set $c_{8|4,4}=0$ and $c_{8|4,2,2}>0$.

To summarize, we have thus shown that to leading order in $v^2$ of the 
translational zero mode, the eighth- and twelfth-order Hamiltonians
are stable and so is the vacuum of course.
We expect the same to hold true for the tenth-order Hamiltonian, but
it is not straightforward to prove it. 

Although the proof of stability in the general case for general
perturbations and to next-to-leading order turns out not to be
straightforward, we would like to make the following conjecture.
\begin{conjecture}
The minimal Hamiltonians of orders 8, 10 and 12, in
Eqs.~\eqref{eq:H8lambda}, \eqref{eq:H10lambda}
and \eqref{eq:H12lambda} are stable to leading order in low-energy
perturbations and in turn so is the vacuum.
\end{conjecture}

If instead we do not expand the Hamiltonians in $v$, but analyze the
conditions for the Hamiltonians to remain positive, we get
\begin{align}
\mathcal{H}_8: \qquad
&1 - 4(w_i^2 + w_j^2) \geq 0, \qquad {\rm for}\ \ i\neq j,\non
&c_{8|4,2,2}\left(1 - 4w_i^2(w^{\rm T} w)\right) - 8 c_{8|4,4} w_j^2
w_k^2 \geq 0, \qquad {\rm for}\ \ i\neq j\neq k,\non
\mathcal{H}_{10}: \qquad
&1 - 4(w^{\rm T}w - w_i^2 + w_1^2w_2^2 + w_1^2w_3^2 + w_2^2w_3^2 +
w_j^2w_k^2) \geq 0, \qquad {\rm for}\ \ i\neq j\neq k,\non
\mathcal{H}_{12}: \qquad
&1 - 4(w^{\rm T}w + 2w_1^2w_2^2 + 2w_1^2w_3^2 + 2w_2^2w_3^2) \geq 0.
\end{align}
Let us start with $\mathcal{H}_8$; if we choose to set $c_{8|4,4}=0$,
the problem of stability simplifies to
\beq
1 - 4w_i^2(w^{\rm T} w) \geq 0,
\eeq
which by the parametrization \eqref{eq:wparam} can be written as
\begin{align}
&\frac{1 + 2v^2 - v^4 + 2v^4(\cos^2\theta
  + \cos(2\chi)\sin^2\theta)}{(1+v^2)^2} \geq 0, \non
&\frac{1 + 2v^2 - v^4 + 2v^2\cos(2\theta)}{(1+v^2)^2} \geq 0,
\end{align}
If we take the approach of assuming $\theta$ and $\chi$ to be worst
possible, meaning that their values will lead to the hardest possible
constraint on $v$, then we get $v<1$, but of course we should not
trust velocities close to 1 with a Galilean boost; therefore,
reinstating the $\gamma$ factor, we get
\beq
v < \frac{1}{\sqrt{2}}.
\eeq
A very rough estimate of the energy scale where the effective theory
breaks down is then $(1+v^2)\Lambda\simeq 1.5\Lambda$, i.e.~about 50\%
above the energy scale of the Skyrmion.

Considering now $\mathcal{H}_{10}$; the constraints for positivity
with the parametrization \eqref{eq:wparam} read
\begin{align}
&\frac{8 - 15v^4 + 4v^2(4 + 5v^2)\cos(2\theta) + v^4(3\cos(4\theta) +
8\cos(2\chi)\sin^4\theta)}{8(1+v^2)^2} \geq 0, \non
&1 - \frac{4\cos^2\theta(v^2 + v^4 +
v^4\cos^2\chi\sin^2\theta)}{(1+v^2)^2}
-\frac{4v^2(1 + v^2 +
2v^2\cos^2\theta)\sin^2\theta\sin^2\chi}{(1+v^2)^2} \non
&\qquad\qquad
-\frac{v^4\sin^4\theta\sin^2(2\chi)}{(1+v^2)^2} \geq 0, \non
&1 - \frac{4v^4\cos^2\chi\sin^2\theta}{1+v^2}
-\frac{4\cos^2\theta(v^2 + v^4 +
2v^4\cos^2\chi\sin^2\theta)}{(1+v^2)^2} \non
&\qquad\qquad
-\frac{v^4(\sin^2(2\theta)\sin^2\chi
+ \sin^4\theta\sin^2(2\chi))}{(1+v^2)^2} \geq 0, 
\end{align}
Taking again the approach of minimizing each constraint with respect
to $\theta$ and $\chi$ to get the hardest constraints on $v$, we
arrive at
\begin{align}
&\frac{1 - 2v^2 - 5v^4}{(1+v^2)^2} \geq 0,\non
&\frac{1 - 2v^2 - 3v^4}{(1+v^2)^2} \geq 0,
\end{align}
of which the first one gives the hardest constraint on $v$.
Reinstating the relativistic $\gamma$ factor, we get
\beq
v < \sqrt{\frac{\sqrt{6}}{2} - 1} \simeq 0.474.
\eeq

Considering finally $\mathcal{H}_{12}$; the constraints for positivity 
with the parametrization \eqref{eq:wparam} read
\begin{align}
\frac{8 - 16v^2 - 35v^4 + v^4(4\cos(2\theta) + 7\cos(4\theta) +
8\cos(4\chi)\sin^4\theta)}{8(1+v^2)^2} \geq 0,
\end{align}
whose hardest constraint on $v$ is
\beq
\frac{2 - 4v^2 - 22v^4}{2(1+v^2)^2} \geq 0.
\eeq
Reinstating the relativistic $\gamma$ factor, we get the constraint
\beq
v < \sqrt{\frac{\sqrt{26} - 4}{5}} \simeq 0.469.
\eeq

We note that increasing the order of the Lagrangian, slightly reduces
the maximal scale at which the theory will break down.
This is somewhat counter intuitive, but we should recall that we work
at a fixed order of derivatives in the $i$-th direction and increase
the total number of derivatives.

We have thus shown that relativistic speeds of the order of
about half the speed of light are necessary before the effective
theory will break down.

\section{Conclusion and discussion}\label{sec:discussion}

In this paper, we have constructed a formalism for higher-derivative
theories based on O(4) invariants. We started with reviewing the
construction made by Marleau, but found that it possesses an
instability in the static energy for all the Lagrangians of higher
than sixth order in derivatives.
The instability can be triggered by a baby-Skyrmion string-like
perturbation that can then run away (see App.~\ref{app:runaway}). 
The problem of the latter construction is the desire to limit the
radial profile function to have a second-order radial equation of
motion. This comes at the cost of the angular derivatives conspiring
at large order in derivatives to create negative terms. This can be
seen by writing the static Lagrangian in terms of eigenvalues of the
derivatives of the O(4) invariants.
We cure the instability by constructing an isotropic construction
where no special direction (e.g.~radial) is preferred to have lower
order in derivatives than others.
This construction necessitates four derivatives in the $i$-th
direction for the Lagrangians of order 8, 10 and 12.

We successfully constructed positive definite static energies for the
Lagrangians of order 8, 10 and 12 with very simple interpretations.
The eighth-order Lagrangian can be interpreted as the Skyrme term
squared plus the Dirichlet energy (normal kinetic term) multiplied by
the BPS-Skyrme term.
The tenth-order Lagrangian instead can be interpreted as the Skyrme
term multiplied by the BPS-Skyrme term.
Finally, the twelfth-order Lagrangian can simply be understood as the
BPS-Skyrme term squared. 

Although our construction straightforwardly yields stable static
energies for higher-order systems, constructing the full Hamiltonians
revealed that time-dependent perturbations may potentially destabilize
the system and in turn its solitons.
The (dynamical) instability we found is intrinsically
different from the 
Ostrogradsky instability as it is not related to the Hamiltonian phase
space being enlarged, but just to the canonical momenta being
nonlinear and in turn inducing terms of both signs.
The nonlinearity induces two effects that destabilize the Hamiltonian;
one is simply the square of the inverse metric for four time
derivatives, which remains negative. The other effect is that 
under the Legendre transform from the Lagrangian to the Hamiltonian,
nonlinearities break the normal would-be SO(4) symmetry (which is
basically the Wick rotated SO(3,1) Lorentz symmetry).
Although this may not be problematic itself, it induces terms of both
signs in our construction.
After reducing the expressions using the eigenvalue formalism, we
obtain clear-cut conditions for positivity of the Hamiltonian given in
terms of one of the vectors of the diagonalization matrix, which has
the physical interpretation of a rotation of the strain tensor into
the time-direction. 
Further analysis may reveal whether this effect truly destabilizes the
Hamiltonian or not.

Finally, we argued that to leading order in time-dependence of the
perturbations under consideration, our construction is stable.
We checked this to leading order in velocity showing that the
Hamiltonians of eighth and twelfth order do not destabilize.
In case of the tenth-order Hamiltonian, we have not been able to prove
stability to leading order although we expect the same to hold true
also in this case.
We conjectured that the Hamiltonians corresponding to the minimal
Lagrangians in our construction are stable to leading orders of
general low-energy perturbations and in turn so is the vacuum. 

It will be interesting to study the dynamical instability
that we encountered here more in detail and to see whether it is
possible to cure it. 
One hope could be to use only odd powers of squared time derivatives,
giving always an odd number of inverse metric factors.
This may, however, not be enough to construct a manifestly positive
Hamiltonian due to the second instability effect that we discussed
above.

Although it may be less likely in our case, it is possible that
dynamical stability can be achieved in some parts of parameter space,
i.e.~for certain values of the constants in the Lagrangians.
For a simpler dynamical system, namely the Pais-Uhlenbeck
oscillator \cite{Pais:1950za} islands of stability were found for
several interacting systems and even bounded Hamiltonians can be found
in some
cases \cite{Smilga:2005gb,Smilga:2008pr,Pavsic:2013noa,Pavsic:2016ykq}.
In our Lagrangians it seems less likely to be possible, because the
instability that we found also manifests itself with just a single
overall coupling constant that can be scaled away.

Another hope for a manifestly stable Hamiltonian could be some
construction with infinitely many derivative terms resummed in a
clever fashion.\footnote{
In Ref.~\cite{Eto:2005cc}, the Skyrme model was constructed 
as the low-energy effective theory on a domain wall 
up to the fourth-derivative order \cite{Eto:2005cc}. 
However, a non-Skyrme term containing 
four time derivatives also exists at this order 
\cite{Eto:2012qda}. 
The effective Lagrangian looks unstable at this order, 
but the domain wall itself must be stable for a topological reason. 
Probably, all terms with infinitely many derivatives 
cure this problem.
} 

One of our motivations to construct a higher-order Skyrme-like term
was to probe whether black hole Skyrme hair is stable only for the
Skyrme term or unstable only for the BPS-Skyrme term
\cite{Gudnason:2015dca,Gudnason:2016kuu,Adam:2016vzf,Perapechka:2016cof}. 

In our construction with minimal number of derivatives in the $i$-th
direction -- which we call the minimal Lagrangians -- all time
derivatives are necessarily multiplied by spatial
derivatives to leading order. Therefore if some instability occurs, it
will be amplified by the presence of solitons. 

Our higher-order terms if added to the Skyrme model will give
corrections to the properties of the Skyrmions, including the mass,
size and binding energy.
Not only Skyrmions, but also the
Skyrme-instanton \cite{Speight:2007ax,Gudnason:2016iex} will receive
corrections from the new higher-order terms.
It will be interesting to study such corrections in the future. 

Another interesting direction will be a supersymmetric extension of
our discussion. 
While supersymmetric extensions of the Skyrme model (of the fourth
order) were studied in
Refs.~\cite{Bergshoeff:1984wb,Freyhult:2003zb,Queiruga:2015xka,Gudnason:2015ryh},
a discussion of topological solitons in supersymmetric theories with
more general higher-derivative terms can be found in
e.g.~Refs.~\cite{Adam:2011hj,Adam:2013awa,Nitta:2014pwa,Bolognesi:2014ova,Nitta:2014fca,Nitta:2015uba}.

\subsection*{Acknowledgments}

We thank Martin Speight for useful discussions. 
S.~B.~G.~thanks the Recruitment Program of High-end Foreign
Experts for support.
The work of S.~B.~G.~was supported by the National Natural Science
Foundation of China (Grant No.~11675223).
The work of M.~N.~is supported in part by a Grant-in-Aid for
Scientific Research on Innovative Areas ``Topological Materials
Science'' (KAKENHI Grant No.~15H05855) from the the Ministry of Education,
Culture, Sports, Science (MEXT) of Japan,
by the Japan Society for the Promotion of Science
(JSPS) Grant-in-Aid for Scientific Research (KAKENHI Grant
No.~16H03984) and by the MEXT-Supported Program for the Strategic
Research Foundation at Private Universities ``Topological Science''
(Grant No.~S1511006).

\appendix

\section{Baby-Skyrmion string run-away perturbation in the Marleau
construction}\label{app:runaway}

In this appendix, we will provide an example of the instability for
illustrative purposes.
Let us for concreteness limit the example to a system with a kinetic
term and an eighth-order Marleau Lagrangian of
Eq.~\eqref{eq:Marleau8simpl}, 
\begin{align}
\mathcal{L} &= \mathcal{L}_2 + \mathcal{L}_8^{\rm Marleau} \non
&= -\ab{1} - \ab{3}\ab{1} + \frac{3}{16}\ab{2}^2
  + \frac{9}{8}\ab{2}\ab{1}^2 - \frac{5}{16}\ab{1}^4,
\end{align}
where we have set $a_1=a_{3,1}=1$ for simplicity (since there are only
two constants, they correspond just to setting the length and energy
units). 

Instead of evolving the full equations of motion, let us just make a
simplified simulation, i.e.~cooling the static equations of motion.
That system can be written as
\begin{align}
-n_{ij}^b\sum_{r=1}^{3}\frac{\p^2\ab{r}}{\p n_i^a\p n_j^b}
  \frac{\p\mathcal{L}}{\p\ab{r}}
-n_{ij}^b\sum_{r,s=1}^{3}\frac{\p\ab{r}}{\p n_i^a} \frac{\p\ab{s}}{\p n_j^b}
  \frac{\p^2\mathcal{L}}{\p\ab{r}\p\ab{s}} = n_0^a.
\end{align}

For illustrative purposes, we will choose a 1-Skyrmion and perturb the
tale of it with a baby-Skyrmion string. The baby-Skyrmion string
carries no baryon number and in the normal Skyrme model it will just
be some extra energy that can be radiated away to find just the
1-Skyrmion being the minimum of the energy.

In this example, however, we have switched the Skyrme term for the
eighth-order Marleau term and hence as shown in
Eq.~\eqref{eq:baby-Skyrme-instability}, the baby-Skyrmion string will 
give rise to a negative energy density that can cause a run-away.

\begin{figure}[!htp]
\begin{center}
\includegraphics[width=0.4\linewidth]{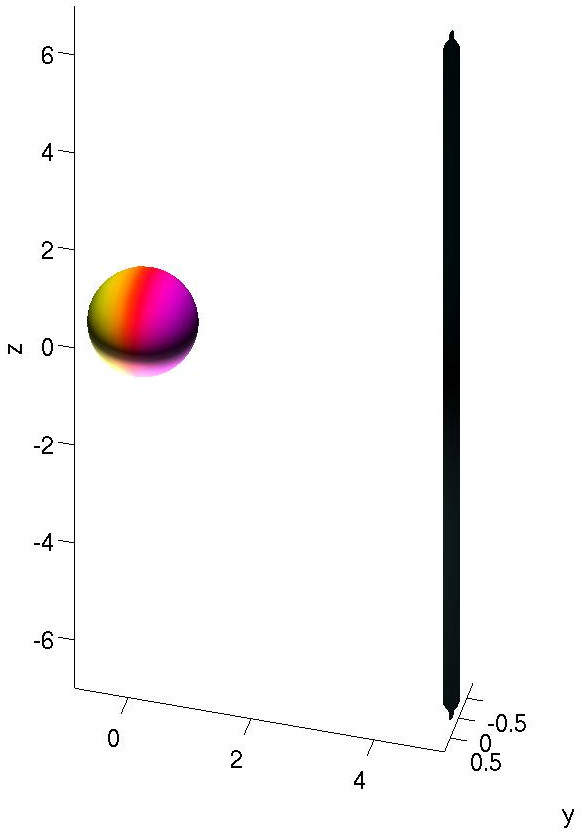}
\caption{The absolute value of the energy density $|\mathcal{E}|$ of
the configuration containing a 1-Skyrmion (the colored sphere) and a
baby-Skyrmion string (the black vertical string). }
\label{fig:En0_100s7_3d}
\end{center}
\end{figure}

\begin{figure}[!htp]
\begin{center}
\includegraphics[width=0.3\linewidth]{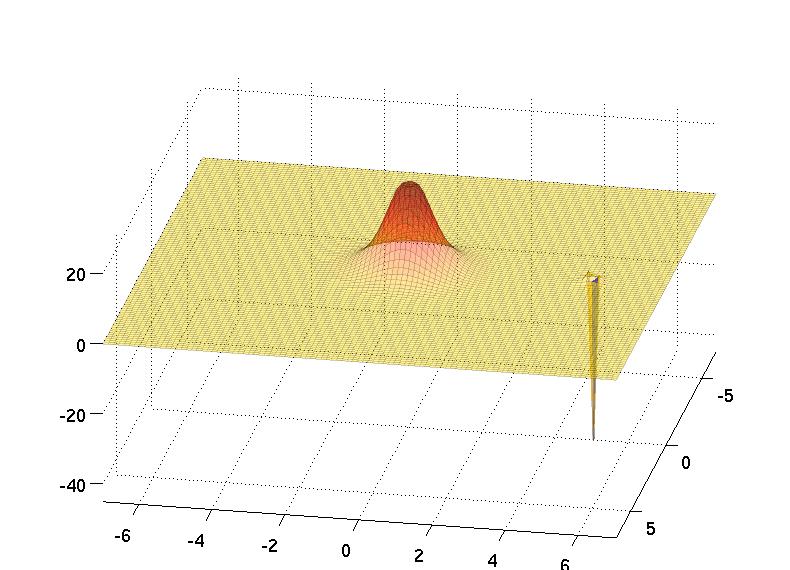}
\includegraphics[width=0.3\linewidth]{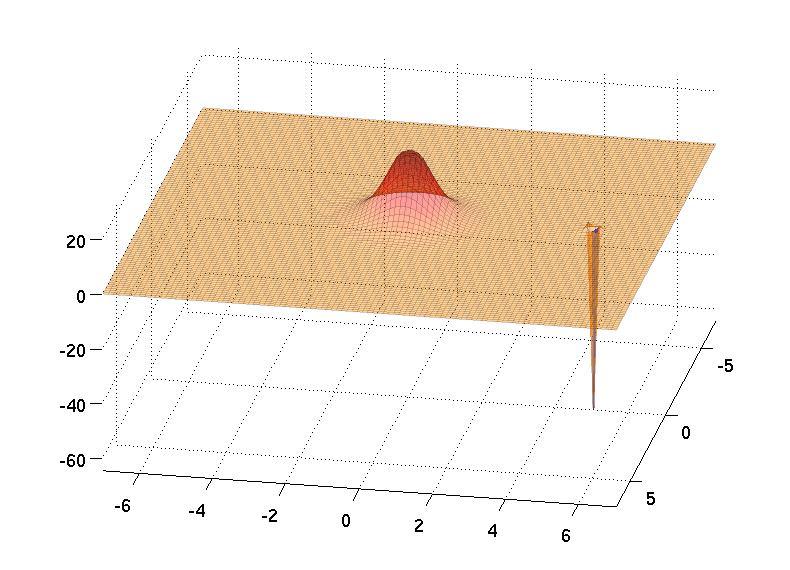}
\includegraphics[width=0.3\linewidth]{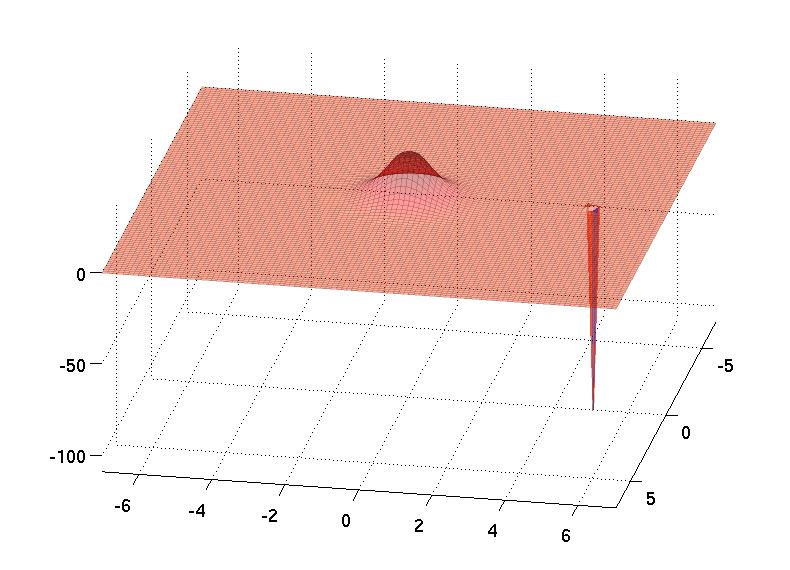}
\caption{Cooling of the configuration shown in
Fig.~\ref{fig:En0_100s7_3d} at time $\tau=0$, $\tau=60$ and
$\tau=120$, respectively. The figure shows the energy density in an
$xy$-slice at fixed $z=0$.
It is seen from the figure that the 1-Skyrmion is unchanged as cooling
time increases, but the energy of the baby-Skyrmion string (to the
right) is growing negative. } 
\label{fig:En0_100s7}
\end{center}
\end{figure}

\begin{figure}[!htp]
\begin{center}
\includegraphics[width=0.6\linewidth]{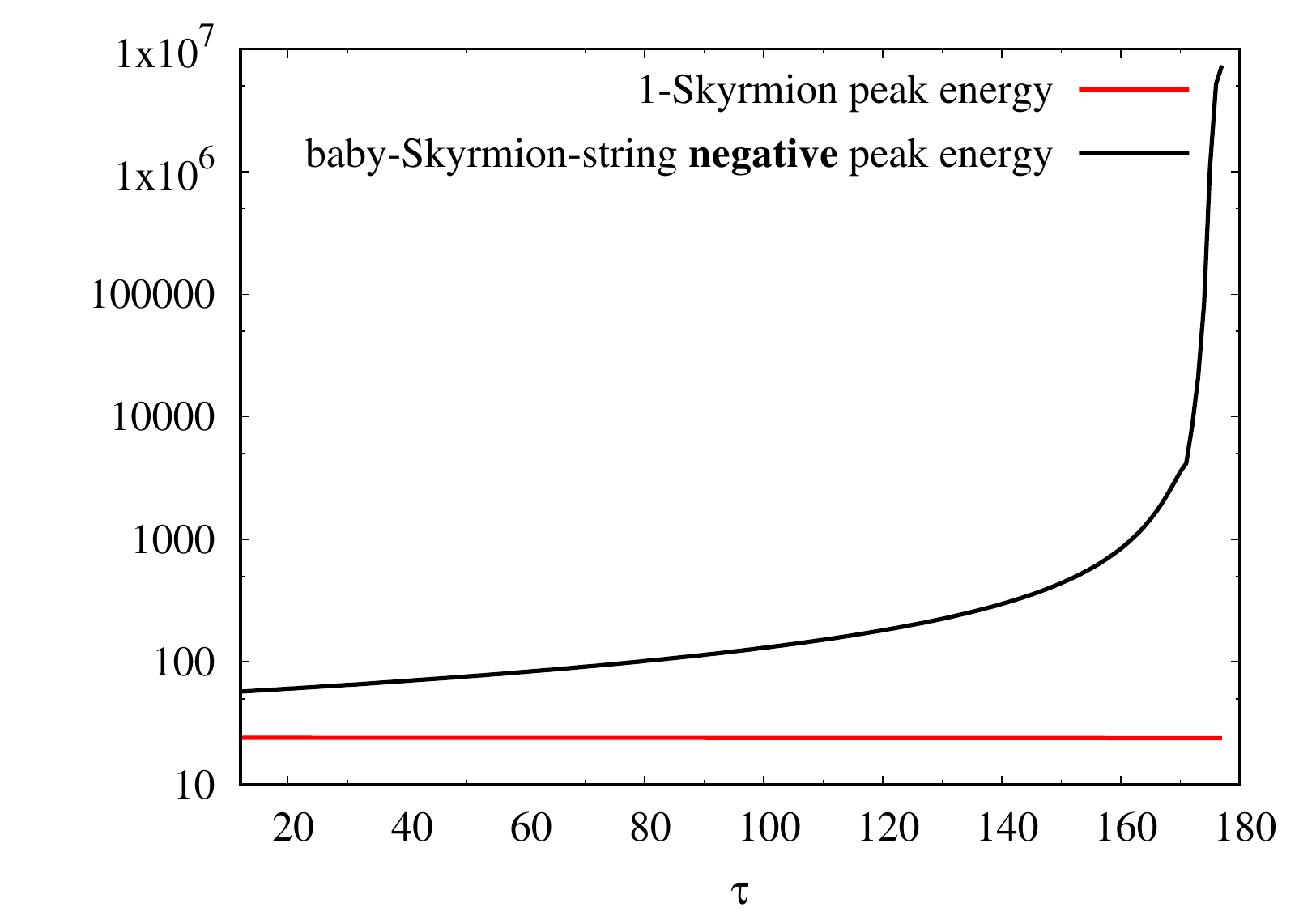}
\caption{Peak energies: positive for the 1-Skyrmion and negative for
the baby-Skyrmion string as functions of the cooling time $\tau$.
It is seen from the figure that the 1-Skyrmion is unchanged, but the
energy of the baby-Skyrmion string (to the right) is growing
negative. }
\label{fig:Ens100s7}
\end{center}
\end{figure}

In Fig.~\ref{fig:En0_100s7_3d} is shown the configuration at the
initial time. The 1-Skyrmion is already the minimum of the energy
functional and its fields have been found using the hedgehog
Lagrangian \eqref{eq:L2nf} with $n=4$. The baby-Skyrmion string is not
a topological object, but just a perturbation added to the
configuration.
In Fig.~\ref{fig:En0_100s7} is shown a series of three snapshots in
cooling time $\tau=0,60,120$ of the configuration. The 1-Skyrmion is
stable and remains a solution, but the baby-Skyrmion string is seen to
grow more and more negative.
Finally, we show the peak energies of the two objects in
Fig.~\ref{fig:Ens100s7}. The 1-Skyrmion has positive peak energy that
remains stable, whereas the baby-Skyrmion string has
a \emph{negative} peak energy that grows rapidly more and more
negative. 
This nicely illustrates the baby-Skyrmion string instability found in
the Marleau construction.

\section{Difference from the Ostrogradsky
instability}\label{app:diffOstrogradsky}

Let us compare the simplest possible term giving rise to four time
derivatives in our theories with that of the Ostrogradsky-like
theories, 
\begin{align}
\mathcal{L} &= -\ab{1}^2 \non
&= -(\p_\mu\mathbf{n}\cdot\p^\mu\mathbf{n})^2
\label{eq:Lsimpl}
\end{align}
which is just the kinetic term squared.
Recall that the nonlinear sigma model constraint
$\mathbf{n}\cdot\mathbf{n}=1$ implies
\beq
\frac{1}{2}\p_\mu(\mathbf{n}\cdot\mathbf{n}) =
\mathbf{n}\cdot\p_\mu\mathbf{n} = 0.
\label{eq:constraint}
\eeq
Consider now integrating the Lagrangian \eqref{eq:Lsimpl} by parts as
\begin{align}
\mathcal{L} &= -(\p_\mu\mathbf{n}\cdot\p^\mu\mathbf{n})
(\p_\nu\mathbf{n}\cdot\p^\nu\mathbf{n}) \non
&= -\p_\mu\left[(\mathbf{n}\cdot\p^\mu\mathbf{n})
  (\p_\nu\mathbf{n}\cdot\p^\nu\mathbf{n})\right]
+ (\mathbf{n}\cdot\p_\mu\p^\mu\mathbf{n})
  (\p_\nu\mathbf{n}\cdot\p^\nu\mathbf{n})
+ (\mathbf{n}\cdot\p^\mu\mathbf{n})
  \p_\mu(\p_\nu\mathbf{n}\cdot\p^\nu\mathbf{n}) \non
&= (\mathbf{n}\cdot\p_\mu\p^\mu\mathbf{n})
  (\p_\nu\mathbf{n}\cdot\p^\nu\mathbf{n}) \non
&= \p_\nu\left[(\mathbf{n}\cdot\p_\mu\p^\mu\mathbf{n})
  (\mathbf{n}\cdot\p^\nu\mathbf{n})\right]
-\left[\p_\nu(\mathbf{n}\cdot\p_\mu\p^\mu\mathbf{n})\right]
  (\mathbf{n}\cdot\p^\nu\mathbf{n})
-(\mathbf{n}\cdot\p_\mu\p^\mu\mathbf{n})
  (\mathbf{n}\cdot\p_\nu\p^\nu\mathbf{n}) \non
&= -(\mathbf{n}\cdot\p_\mu\p^\mu\mathbf{n})
  (\mathbf{n}\cdot\p_\nu\p^\nu\mathbf{n}),
\label{eq:int_by_parts}
\end{align}
which obviously differs from the Ostrogradsky-like
Lagrangian \cite{Woodard:2015zca}
\beq
\mathcal{L} &= -\p_\mu\p^\mu\mathbf{n}\cdot\p_\nu\p^\nu\mathbf{n}.
\label{eq:LOstrogradsky_like}
\eeq
The reason why we do not have the Ostrogradsky instability, exactly,
is because the propagator of Eq.~\eqref{eq:int_by_parts} is not $p^4$; 
it remains $p^2$ and the term is still just a product of two kinetic
terms. 

Trying to formally manipulate the Ostrogradsky-like
Lagrangian \eqref{eq:LOstrogradsky_like}, we get
\begin{align}
\mathcal{L} 
&= -\p_\mu\p^\mu\mathbf{n}\cdot\p_\nu\p^\nu\mathbf{n} \non
&= -(\p_\mu\p^\mu\mathbf{n}\cdot\p_\nu\p^\nu\mathbf{n})
  (\mathbf{n}\cdot\mathbf{n}) \non
&= -\p_\mu\left[(\p^\mu\mathbf{n}\cdot\p_\nu\p^\nu\mathbf{n})
  (\mathbf{n}\cdot\mathbf{n})\right]
+(\p^\mu\mathbf{n}\cdot\p_\mu\p_\nu\p^\nu\mathbf{n})
  (\mathbf{n}\cdot\mathbf{n})
+2(\p^\mu\mathbf{n}\cdot\p_\nu\p^\nu\mathbf{n})
  (\mathbf{n}\cdot\p_\mu\mathbf{n}) \non
&= -\p_\mu\left[(\p^\mu\mathbf{n}\cdot\p_\nu\p^\nu\mathbf{n})\right]
+(\p^\mu\mathbf{n}\cdot\p_\mu\p_\nu\p^\nu\mathbf{n})
+2\p_\nu\left[(\p^\mu\mathbf{n}\cdot\p^\nu\mathbf{n})
  (\mathbf{n}\cdot\p_\mu\mathbf{n})\right] \non
&\phantom{=\ }
-2(\p^\mu\p_\nu\mathbf{n}\cdot\p^\nu\mathbf{n})
  (\mathbf{n}\cdot\p_\mu\mathbf{n})
-2(\p^\mu\mathbf{n}\cdot\p^\nu\mathbf{n})
  (\mathbf{n}\cdot\p_\mu\p_\nu\mathbf{n})
-2(\p^\mu\mathbf{n}\cdot\p^\nu\mathbf{n})
  (\p_\nu\mathbf{n}\cdot\p_\mu\mathbf{n}) \non
&= -\p_\mu\left[(\p^\mu\mathbf{n}\cdot\p_\nu\p^\nu\mathbf{n})\right]
+(\p^\mu\mathbf{n}\cdot\p_\mu\p_\nu\p^\nu\mathbf{n})
+2\p_\nu\left[(\p^\mu\mathbf{n}\cdot\p^\nu\mathbf{n})
  (\mathbf{n}\cdot\p_\mu\mathbf{n})\right]
\non&\phantom{=\ }
-\p^\mu\left[(\p_\nu\mathbf{n}\cdot\p^\nu\mathbf{n})
  (\mathbf{n}\cdot\p_\mu\mathbf{n})\right]
+ (\p_\nu\mathbf{n}\cdot\p^\nu\mathbf{n})
  (\p_\mu\mathbf{n}\cdot\p^\mu\mathbf{n})
+ (\p_\nu\mathbf{n}\cdot\p^\nu\mathbf{n})
  (\mathbf{n}\cdot\p_\mu\p^\mu\mathbf{n})
\non&\phantom{=\ }
-2(\p^\mu\mathbf{n}\cdot\p^\nu\mathbf{n})
  (\mathbf{n}\cdot\p_\mu\p_\nu\mathbf{n})
-2(\p^\mu\mathbf{n}\cdot\p^\nu\mathbf{n})
  (\p_\nu\mathbf{n}\cdot\p_\mu\mathbf{n}) \non
&= -\p_\mu\left[(\p^\mu\mathbf{n}\cdot\p_\nu\p^\nu\mathbf{n})\right]
+(\p^\mu\mathbf{n}\cdot\p_\mu\p_\nu\p^\nu\mathbf{n})
+ (\p_\nu\mathbf{n}\cdot\p^\nu\mathbf{n})
  (\p_\mu\mathbf{n}\cdot\p^\mu\mathbf{n})
\non&\phantom{=\ }
+ (\p_\nu\mathbf{n}\cdot\p^\nu\mathbf{n})
  (\mathbf{n}\cdot\p_\mu\p^\mu\mathbf{n})
-2(\p^\mu\mathbf{n}\cdot\p^\nu\mathbf{n})
  (\mathbf{n}\cdot\p_\mu\p_\nu\mathbf{n})
-2(\p^\mu\mathbf{n}\cdot\p^\nu\mathbf{n})
  (\p_\nu\mathbf{n}\cdot\p_\mu\mathbf{n}),
\label{eq:LfromO}
\end{align}
where in the last equation we have used Eq.~\eqref{eq:constraint}.
We can identify the $\ab{1}^2$ and the $-2\ab{2}$ terms in the last
equation.
However, deriving the constraint \eqref{eq:constraint} once more, we
get that
\beq
\p_\nu\mathbf{n}\cdot\p_\mu\mathbf{n}
+ \mathbf{n}\cdot\p_\mu\p_\nu\mathbf{n} = 0,
\eeq
where $\nu$ can also be equal to $\mu$ and summed over; it is a
general statement derived from the nonlinear sigma model constraint.
Using this relation, however, we can simplify Eq.~\eqref{eq:LfromO} to
\begin{align}
\mathcal{L}
&= -\p_\mu\left[(\p^\mu\mathbf{n}\cdot\p_\nu\p^\nu\mathbf{n})\right]
+(\p^\mu\mathbf{n}\cdot\p_\mu\p_\nu\p^\nu\mathbf{n}),
\end{align}
which is simply the Ostrogradsky-like Lagrangian that we started with.
Therefore, we can see that the Lagrangian \eqref{eq:Lsimpl}
is \emph{not} just the Ostrogradsky-like
Lagrangian \eqref{eq:LOstrogradsky_like} up to a total derivative.

Nevertheless, this exercise should show that the Ostrogradsky system
is intrinsically different from our Lagrangians and that we do not
have $p^4$ in the propagator, but just a highly nonlinear theory.

\end{document}